\documentclass[a4paper,11pt]{article}
\pdfoutput=1 

\usepackage[dvipsnames]{xcolor}
\usepackage{graphicx,color}
\usepackage{jheppub} 
\usepackage{enumitem}
\usepackage{braket}
\usepackage{slashed}
\usepackage[utf8]{inputenc}
\usepackage{hyperref}
\usepackage[most]{tcolorbox}
\usepackage{float}
\usepackage{caption}
\usepackage{subcaption}
\usepackage{makecell}
\usepackage{epsfig}
\usepackage{amsmath,amssymb}
\usepackage{hyperref}
\usepackage{ulem}
\usepackage{tikz}
\usepackage{comment}
\usetikzlibrary{shapes.geometric, arrows}

\newcommand{\tti}[1]{\text{\tiny{#1}}}

\hypersetup{
	colorlinks=true,
	citecolor=black,
	filecolor=black,
	linkcolor=blue,
	urlcolor=blue
}

\newcommand{\lr}[1]{\left(#1\right)}

\title{Gravitational Waves from High Temperature Strings}

\author[a]{Andrew R. Frey,}
\author[a]{Ratul Mahanta,}
\author[b]{Anshuman Maharana,}
\author[c,d,e]{Fernando Quevedo,}
\author[c]{Gonzalo Villa}

\affiliation[a]{\small Department of Physics and Winnipeg Institute for Theoretical Physics,
University of Winnipeg, 515 Portage Avenue, Winnipeg, Manitoba R3B 2E9, Canada}
\affiliation[b]{\small Harish Chandra Research Institute, A CI of Homi Bhabha National Institute,
Chattnag Road, Jhunsi, Prayagraj (Allahabad) - 211019, India}
\affiliation[c]{\small DAMTP,    University of Cambridge, Wilberforce Road,  Cambridge, CB3 0WA, UK
}
\affiliation[d]{CERN, Theoretical Physics Department, Geneve 23, Switzerland}
\affiliation[e]{New York University Abu Dhabi, PO Box 128199, Saadiyat Island, Abu Dhabi, UAE}
 
\emailAdd{a.frey@uwinnipeg.ca}
\emailAdd{r.mahanta@uwinnipeg.ca}
\emailAdd{anshumanmaharana@hri.res.in}
\emailAdd{fq201@damtp.cam.ac.uk}
\emailAdd{gv297@cam.ac.uk}

\abstract{ We study  finite temperature effects in string cosmology and their potential gravitational wave signature. Expanding on our recent work \cite{Frey:2023khe}, we consider a general configuration of highly excited open and closed strings at high enough temperature to  be in the Hagedorn phase in 3+1 dimensions, in order to explore its cosmological implications. We find conditions, which can be satisfied in compactifications with moduli stabilization, that allow the long strings to remain in equilibrium in a controlled effective field theory, with equilibration driven by the joining and splitting of the dominant open string population.
We calculate the emission rate of gravitons by long open strings, which we show is determined by ten dimensional flat space transition amplitudes available
in the literature, and then find the total gravitational wave spectrum generated by the gas of long strings.
The gravitational wave spectrum has robust characteristics.
It peaks at frequencies of order 50-100 GHz, the same as for gravitational waves from the reheating epoch of the Standard Model. But the amplitude
of the string signal is {\it significantly larger} than predicted by the Standard Model and its field theoretic
extensions. The amplitude and other physical observables (such as the contribution to $\Delta N_{\text eff}$) are directly proportional to the string scale 
 $M_s$; indicating that a potential signal may also determine the string scale. Our calculations provide one of the few examples of a signal of stringy origin that dominates over the field theory predictions. We give a physical explanation of our results and discuss further implications.    

}
\preprint{CERN-TH-2024-108}
\begin{document}
\maketitle
\flushbottom

\section{Introduction}
String theory is above all a fundamental theory of gravity. In the long road towards searching for potential experimental signatures of string theory, gravitational waves (GW) stand out as arguably  the most relevant prospect for model independent tests of the theory. The impressive progress on the detection of gravitational waves during the past decade makes it hopeful that eventually gravitational waves predicted from a fundamental theory could be discovered in the not-too-distant future.

There are many potential sources of stochastic gravitational wave backgrounds from physics beyond the standard model (BSM), from cosmic strings to reheating, non-topological defects, etc (see~\cite{Caprini:2018mtu,Aggarwal:2020olq,Roshan:2024qnv} for reviews). Most of them can be incorporated into string theoretical frameworks. However it is important to try to study sources of gravitational waves that are intrinsically stringy in nature.\\

A general, model independent property of string theory is the existence of the Hagedorn temperature, which hints at a stringy phase that could be realised at the very early universe. This is a consequence of the exponentially growing number of massive string states.
Depending on their characteristics, systems approaching the Hagedorn temperature~\cite{Hagedorn:1965st,Hagedorn:1967tlw,Hagedorn:1967dia,Frautschi:1971ij,Carlitz:1972uf,Cabibbo:1975ig}  may undergo a phase transition (see e.g.~\cite{Sathiapalan:1986db, Kogan, Atick:1988si,Mitchell:1987hr,Mitchell:1987th,Deo:1989bv,Abel:1999rq,Barbon:2004dd,Copeland:1998na}) or instead slowly approach a constant limiting, or ultimate temperature (see e.g.~\cite{Huang:1970iq, Sundborg:1984uk, Bowick:1985az, Tye:1985jv, Giddings:1989xe}), with the entropy injected being used to populate this exponentially large number of states\footnote{In string theory, this holds until the energy density is sufficient to nucleate brane-antibrane pairs and the description breaks down.}. These two different behaviours are usually called non-limiting and limiting behaviours, respectively~\cite{Abel:1999rq}. We refer to~\cite{Barbon:2004dd} for a review in these and related issues in string thermodynamics.\\

In the context of string theory, the limiting class of systems feature a \textit{Hagedorn phase} with a gas of highly excited string states dominating the energy density.
This is continuously connected to a low energy radiation phase like the supercritical fluid of water above the critical point is connected to the liquid or gas phases.
The study of this Hagedorn phase in cosmology and some of its potentially observable signatures from {\it gravitational waves} is the subject of this article.\\

A common concern about string thermodynamics is that, because string theory is a theory of gravity, gravitational backreaction necessarily prevents the existence of a static, homogeneous state of thermal equilibrium. 
Based on the out-of-equilibrium dynamics studied in \cite{Frey:2023khe}, we will find conditions for a Hagedorn phase of strings to equilibrate more quickly than the action of gravitational processes (such as Hubble expansion or the Jeans instability). 
The rapid equilibration is driven by splitting and joining interactions of open string endpoints, and it is consistent with effective field theory when there is a large hierarchy between string and Planck scales.
We believe this to be the first consistent description of equilibrium for highly-excited strings in cosmology.
We assume that the Standard Model (SM) is realized in a brane construction and that these long open strings are highly-excited degrees of freedom in the SM sector; that is, the Hagedorn phase is the high-density phase of the SM.\\

To leave an observable imprint, this Hagedorn phase should occur after inflation, and it could indeed have been its endpoint~\cite{Frey:2005jk}.
It is reasonable to wonder if inflation in the effective theory can supply an energy density
greater than the string scale after reheating (see~\cite{Cicoli:2023opf} for a recent review in string cosmology with details of inflation in string theory).
If the inflationary Hubble scale is $H_{inf}$, energy considerations require $H_{inf}> M_s^2/M_p$, where the tension of long strings is $\sim M_s^2$ and set by the local string scale $M_s$.
Meanwhile, remaining in the effective theory requires $H_{inf}\ll M_{KK}$ for Kaluza-Klein scale\footnote{Or the scale of some other tower of states \cite{Ooguri:2006in}.} $M_{KK}$.
As a result, a Hagedorn phase is compatible with inflationary cosmology whenever $M_{KK}/M_s\gg M_s/M_p$. Also, the quantization of the free string, which underlies the computation of the density of states and so the main thermodynamic properties of the system, is a good approximation in this background when these hierarchies are satisfied.
We present compactification scenarios where these hierarchies hold, so we expect a Hagedorn phase whenever the reheating temperature is high enough.\\

It is also important to remark that some alternatives to inflation also feature stringy ingredients (see~\cite{Brandenberger:2023ver} for a recent review) and a phase featuring highly excited strings is natural from this perspective.
As we will see, our results for the GW spectrum only depend in the last steps of the Hagedorn phase and so are robustly independent of what physics sourced it\footnote{It is important
to distinguish our findings with that of  \cite{Brandenberger:2006xi}. We will be considering direct decay of highly excited open strings to gravitons. This leads to an 
amplitude which scales as
 $M_s/M_p$. It is much larger than the $\left( M_s/M_p \right)^4$ scaled amplitude of the tensor modes fluctuations (akin to the tensor modes in
 inflationary scenarios) in \cite{Brandenberger:2006xi} obtained from equilibrium fluctuations of closed strings. }.\\

Besides the many BSM sources of gravitational waves, there is a general stochastic spectrum of GWs that is present even in the SM due to the early universe plasma interacting with gravity~\cite{Ghiglieri:2015nfa,Ghiglieri:2020mhm, Ringwald:2020ist,Muia:2023wru} 
(see also \cite{Barman:2023ymn}).
This GW background has been studied recently and, tracking the cosmic microwave background (CMB), it peaks at a high frequency of order 80 Giga Hertz (GHz), with its peak amplitude depending linearly in the reheating temperature.
A surprising fact is that unless there exists a rare modification of the cosmological history of the universe, extensions of the SM at a given reheating temperature predict a peak amplitude which is typically smaller than the SM prediction~\cite{Ringwald:2020ist,Muia:2023wru}.\\

It is worth noting in this context that gravitational waves at high frequencies are attracting the attention of theorists and experimentalists (see~\cite{Aggarwal:2020olq} for a review of sources and proposals for experiments).
At the moment of writing there are proposals for GW detection around the GHz band~\cite{Berlin:2021txa}, although none yet achieve sensitivities to stochastic backgrounds of cosmological origin.
It is nevertheless important to clearly identify possible sources --- the GW spectrum from the open string Hagedorn phase turns out to be an interesting target. 
One of the motivations of the present article is to contrast the (B)SM predictions with the string theoretic spectrum arising due to the decay of excited string modes into massless states, including gravitons, and less excited states. The fact that long strings can emit gravitons directly makes their contribution to the GW spectrum dominate over the SM spectrum which comes from scattering with additional gauge coupling suppression. GW emission by SM radiation after the Hagedorn phase is not strong enough to obscure the GW signal of thermal strings. \\

We organize the presentation as follows. Section \ref{sec:thcosmo} deals with
``Hagedorn Cosmology,'' an epoch where the energy density of the universe is
dominated by highly excited (i.e. long) strings in thermal equilibrium. We discuss realistic scenarios that could accommodate a Hagedorn phase of open and closed strings; warped compactifications are natural (but not unique) settings to realise the scenario. We compare the equilibration rates of thermal strings $\Gamma$, computed in~\cite{Frey:2023khe}, to the Hubble scale $H$, during this epoch to gain an understanding of the relevant
reactions and how various species remain in equilibrium and decouple during the cosmological evolution; open string interactions are the key process driving equilibration. We also discuss the validity of our approximations.\\

Section \ref{sec:gw-emission} concerns the emission of gravitons from
the decays of long open strings. We first provide a quantum mechanical analysis which
determines the  form of the coupling of long strings to the graviton in general
backgrounds and see that the matrix element for graviton emission is the same as in a toroidal compactification.
Comparing this
with various graviton emission computations available in the literature (in ten dimensional flat space) we arrive at the general form of the decay rate in the settings of interest.\\ 

In section \ref{sec:sto}, we combine the results of sections \ref{sec:thcosmo} and \ref{sec:gw-emission} to compute the spectral curve of the
stochastic gravitational wave background emitted during an epoch of open string Hagedorn cosmology. Key features of
the spectrum such as the  peak and amplitude are discussed. Also, the corresponding
$\Delta N_{\rm eff}$ (the effective number
of additional neutrino like species at the time of neutrino decoupling)
is computed and compared with observational constraints.\\

The end of this section discusses various phenomenological aspects of the stochastic
background in detail.  We compare
the spectrum  with the spectrum of gravitational waves expected from
the reheating epoch of the Standard Model/BSM models. We consider the most important result is that the GWs spectrum produced by string theory peaks at the same order of magnitude in frequency but it is hierarchically larger than all others.
Our scenario therefore opens up a concrete way to eventually test a key property of string theory. 
For D-brane constructions of the SM, a large hierarchy between string and Planck scales, and sufficiently high-scale inflation, our basic predictions are robust and generic.
Furthermore, we find that the spectrum is proportional to the string scale signaling a way to also determine the string scale. We provide an explanation of why it is expected that the string spectrum is dominant. The reader more interested on the phenomenological aspects may prefer to go directly to this section.\\
  
We discuss future directions and conclude in section \ref{sec:conclude}. We dedicate two appendices to present more details of the scenario we are considering.

\section{String thermodynamics in cosmology}
\label{sec:thcosmo}
In this paper, we consider realistic compactifications with the effect of ingredients that render moduli stabilization and $d=3$ noncompact spatial directions (plus time) which expand due to the backreaction of a homogeneous and isotropic gas of very long open and closed strings (the Hagedorn phase).
In this section we discuss relevant details about equilibrium and out-of-equilibrium notions that need to be tackled in order to study a thermal plasma in an expanding universe.
We begin in~\ref{sec:scenarios} by identifying three possible realistic scenarios which feature a Hagedorn phase, leaving a more systematic discussion of the notions of limiting and non-limiting behaviour in string thermodynamics to Appendix~\ref{sec:setup}.
We then review relevant results of~\cite{Frey:2023khe} that discuss equilibrium configurations and equilibration rates $\Gamma$ of our system of interest in section~\ref{sec:equilibrium}.
This is important because thermal equilibrium in cosmology is an approximate notion which only describes a system provided $\Gamma/H\gg 1$, where $H$ is the Hubble scale induced by the presence of the gas.
This is discussed in sec.~\ref{sec:cosmology}, where we show that the energy density is dominated by highly excited open string degrees of freedom which do maintain equilibrium, and discuss entropy conservation (which allows us to adiabatically track the evolution of the gas).
In addition, we study the validity of our approximations (like neglecting $\alpha'$ corrections in presence of a large energy density) and identify the relevant range of parameter space where the scenario is under control.

\subsection{The Hagedorn phase in realistic scenarios}\label{sec:scenarios}
In the present paper we wish to discuss a thermal system in which the strings propagate in three noncompact spatial directions (and time), which are furthermore worldvolumes of branes.
These assumptions are key for the analysis of the thermodynamics (which we discuss at length in Appendix~\ref{sec:setup}) and it is worth commenting on which realistic scenarios can accommodate them.\\

The idea is that in the thermal gas two new scales appear: first, the length $L$ of the typical string, which grows with the energy of the system and determines the temperature.
Because highly excited strings form random walks, this typical string will spread through another distance scale\footnote{We define length in terms of the mass $M$ of the string as $l= M/M_s^2$ to avoid cumbersome factors of $2\pi$, but note that strictly speaking the length of a string is given by $2\pi M/M_s^2$.} $L_{rms}\equiv \sqrt{L/M_s}$ in each direction.
The thermodynamics of these excited strings depends on how many dimensions of space are large or small compared to $L_{rms}$. \\

If a dimension is noncompact, it is always large comparatively, so the probability for a string to self-intersect (or to intersect a D-brane) depends on the length of the string.
On the other hand, a compact direction with Kaluza-Klein scale $l_{KK}\ll L_{rms}$ is small, and the string fills the whole compact space. In this case, the self-intersection probability is $\sim 1/(l_{KK} M_s)$, independent of the string length.
On the other hand, compact dimensions with $l_{KK}\gg L_{rms}$ are classified as large.\\

\begin{figure} 
\begin{subfigure}{0.6\textwidth}
  \centering  \includegraphics[width=1.0\textwidth]{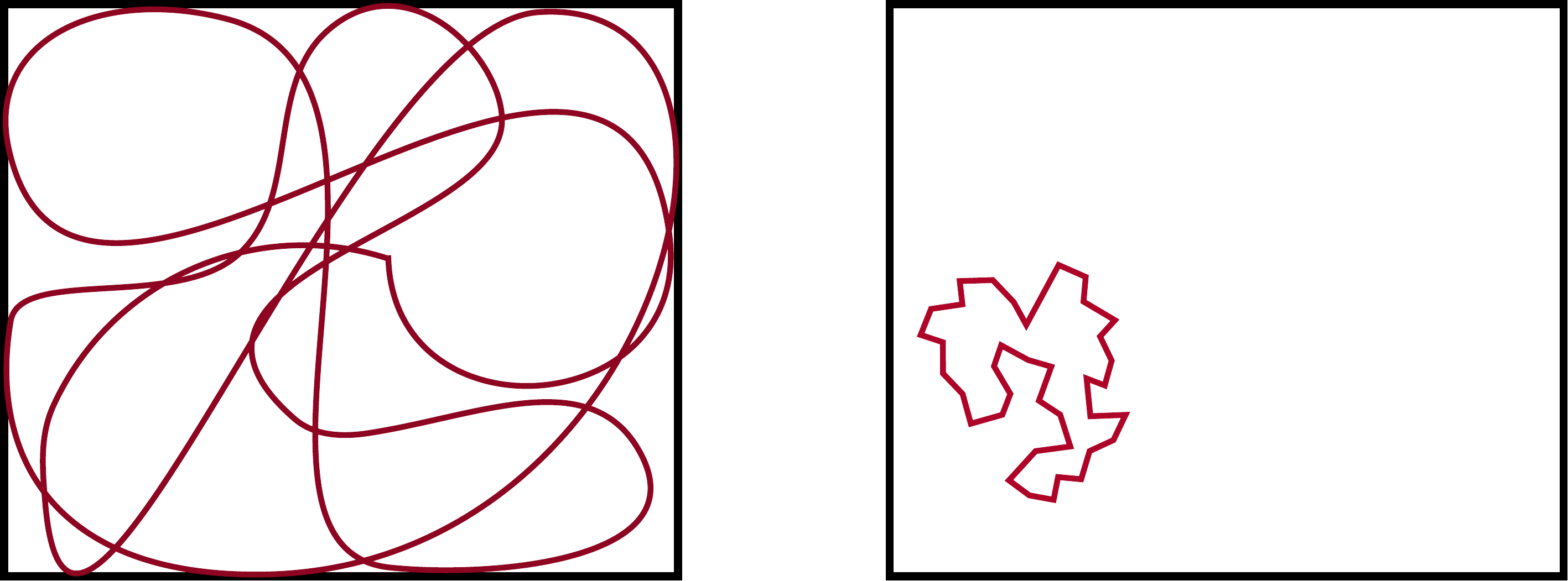}
  \subcaption{Small (left) vs large (right) dimensions}\label{sf:large-small}
\end{subfigure} 
\begin{subfigure}{0.4\textwidth}\centering  \includegraphics[width=0.6\textwidth]{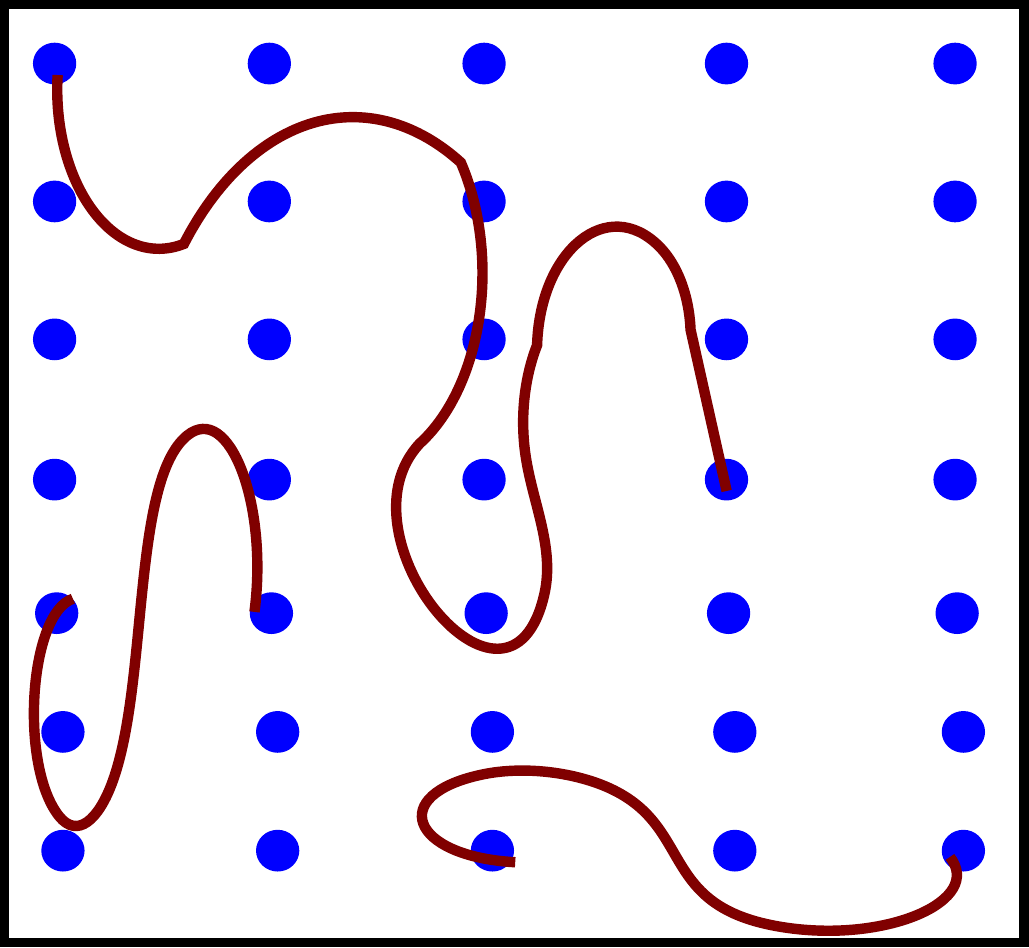}
\subcaption{Small inter-brane separation}\label{sf:gas-branes}
\end{subfigure}
\caption{(\subref{sf:large-small}) Long strings in small dimensions overlap themselves but do not in large dimensions.
(\subref{sf:gas-branes}) Two directions transverse to a gas of parallel D$p$-branes (blue points). The inter-brane separation is small. Red lines represent long strings.}
\label{fig:random-walk}
\end{figure}

In the case of open strings, consider a homogeneous gas of parallel D$p$-branes separated by a length $l_b$ in the transverse directions.
If $l_b$ is comparatively large, the branes are isolated from each other; however, if $l_b\ll L_{rms}$ is small, the long strings can intersect a brane at essentially any point along their length, as if the branes filled the space.
Figure \ref{fig:random-walk} illustrates the comparison of $L_{rms}$ to $l_{KK}$ and $l_b$.
The key point is that if all large dimensions are filled by D-branes (or $l_b$ is small), the thermodynamics of the string gas is well-described by the canonical ensemble, and the temperature approaches the Hagedorn temperature $T_H$ only in the limit of infinite energy density.
We refer the reader to Appendix~\ref{sec:setup} for a more detailed discussion of the role played by large and small directions in the thermodynamics and how to identify them.\\

Here we simply point out three possible scenarios that render all internal dimensions as effectively small (or filled by branes) and the range of validity for which these notions apply. 
The following all have limiting behavior for the thermodynamics:

\begin{itemize}
    \item \textit{A Brandenberger-Vafa scenario with open and closed strings}. 
    The original work of \cite{Brandenberger:1988aj} considered closed strings in 9 compact and small dimensions. We consider D-branes filling 3 noncompact dimensions (and possibly some compact dimensions) with Kaluza-Klein length $l_{KK}$ small in the directions transverse to the branes, which leads to similar thermodynamic behavior.
    For 6 roughly isotropic compact dimensions, we have $M_s/M_p\sim g_s/(M_s l_{KK})^3\ll 1$ when $l_{KK}$ is large in string units.
    For the compact dimensions to be small ($l_{KK}\ll L_{rms}$), the typical string energy is $M_s^2 L \gtrsim M_s^3 l_{KK}^2$.
    While large, we will see that this can still be parametrically less than requirements for the validity of EFT.
    This scenario resembles the left image in figure \ref{sf:large-small}.
    \item \textit{Dense brane scenario}.
    Another possibility is that there is a roughly  homogeneous distribution of parallel branes along all directions in the compact space.
    If that is the case, then the strings only need to be as large as the typical inter-brane separation $l_b < l_{KK} $ and the description would therefore apply at energies lower than the Brandenberger-Vafa case (at higher densities, this becomes a Brandenberger-Vafa scenario).
    This scenario resembles figure \ref{sf:gas-branes}.
    \item \textit{The Jackson-Jones-Polchinski~\cite{Jackson:2004zg} box}.
    This is the most interesting case from the perspective of a realistic compactification.
    As discussed in appendix~\ref{sec:setup}, branes, fluxes and other key ingredients for realistic phenomenology (including the Standard Model and moduli stabilization) will typically backreact on the compact space, rendering a (strongly or not) warped metric. This effect localises the highly excited strings in a string-scale region.
    Importantly, the intersection probability is independent of the length of the highly excited string.
    Then, the internal directions are effectively small, even though the compact dimensions may actually have a large extent.
    In support of this argument, \cite{Canneti:2024iyn} recently considered the single string density of states with spacetime curvature, finding that target space dimensions with worldsheet masses, such as the warped directions, are effectively small, even if they are physically much larger than $L_{rms}$. 
\end{itemize}

\begin{figure} 
  \centering  \includegraphics[width=0.6\linewidth]{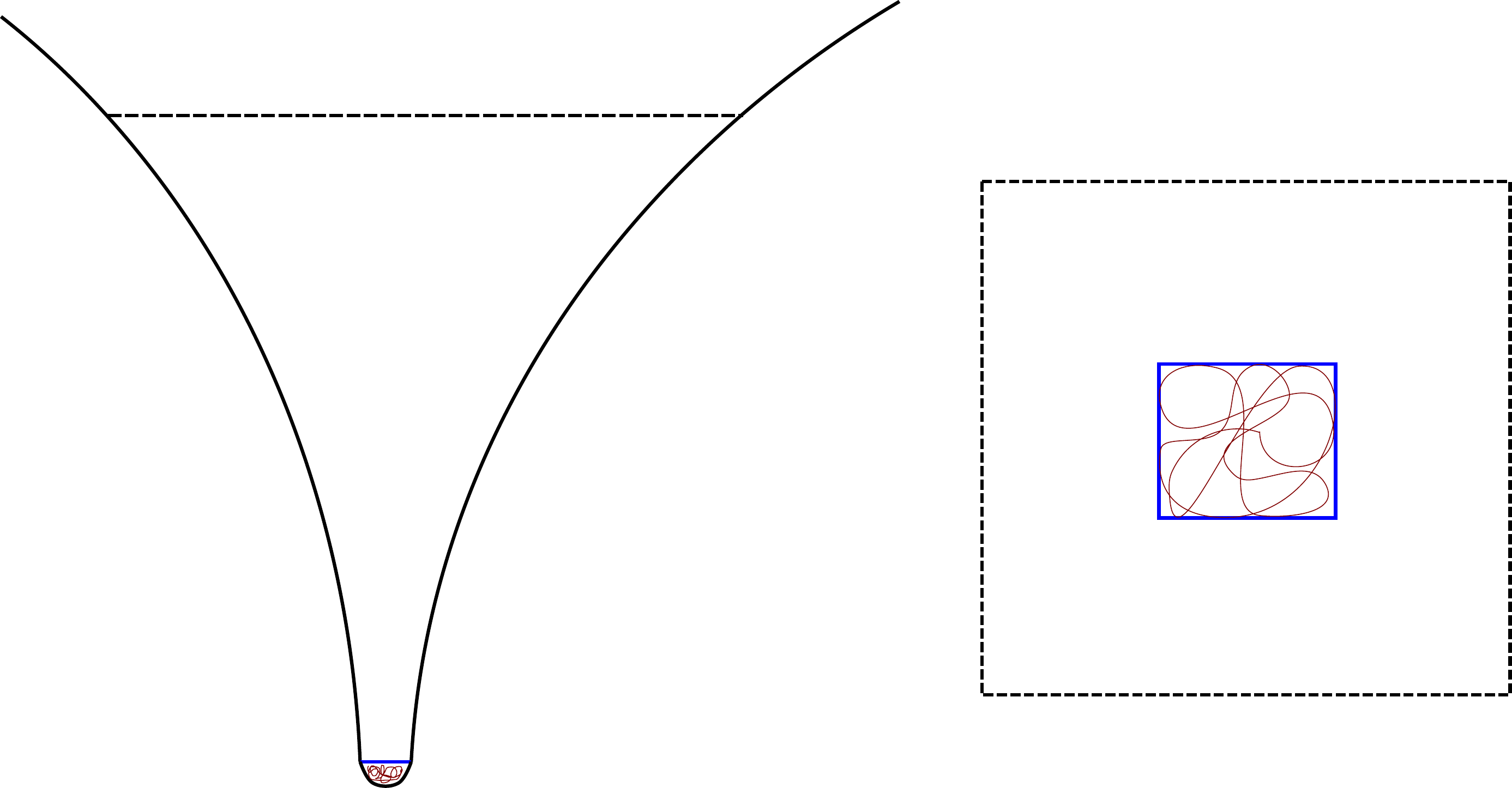}
\caption{A cartoon version of the Jackson-Jones-Polchinski scenario. Strings are confined to a small region at the tip of a throat (blue lines) compared to the full throat (dashed black) or bulk compactification. The right part of the figure is a transverse view. Red lines indicate long strings.}
\label{fig:jjp}
\end{figure}

This confining effect on strings does not require strong warping and can occur for even $\mathcal{O}(1)$ modulation in the warp factor.
Therefore, we will follow the latter scenario, keeping in mind that it is the one that takes into account the effect of generic ingredients of realistic string compactifications.
We review the Jackson-Jones-Polchinski argument~\cite{Jackson:2004zg} in appendix~\ref{sec:setup}\footnote{We also compare the three scenarios in more detail in that appendix.}.
Their conclusion is that highly excited strings get localized in an internal volume of size $\alpha'^3$; we will call this region the ``JJP box.''
We illustrate the JJP box in figure \ref{fig:jjp}.\\

As a result, we approximate the thermodynamics of strings in a warped compactification by strings in flat spacetime with a compactification of volume $\sim \mathcal{O}(1)$ in string units; the long strings fill those compact dimensions.
We can also approximate the warp factor as constant across the compact region (since its curvature should be small in string units).
Then the string thermodynamics are those of strings in three flat noncompact dimensions with spacefilling D-branes; as argued in \cite{Frey:2023khe}, the effect of the compact dimensions is just to modify the interaction coefficients.
Finally, since the 4D metric $g_{\mu\nu}$ appears in the 10D line element as $ds^2 =e^{2A(y)} g_{\mu\nu}dx^\mu dx^\nu+\cdots$, the energy scale of the strings is set by the warped string scale, which we denote $M_s$\footnote{As an example of this effect on string thermodynamics, the Hagedorn temperature in Maldacena-Nu\~{n}ez--Chamseddine-Volkov backgrounds \cite{Maldacena:2000yy,Chamseddine:1997nm} is set by the string scale at the bottom of the throat \cite{PandoZayas:2003jr}.}. 
Note that $M_s$ is simply the higher-dimensional string scale $\alpha'^{-1/2}$ in unwarped scenarios (such as the Brandenberger-Vafa (BV) or dense brane scenarios above) but can still be much smaller than the Planck mass in large volume compactifications.

\subsection{String thermodynamics and equilibration}\label{sec:equilibrium}
Here we review string thermodynamics (at and near equilibrium) in flat spacetime, taking care to restore units. 
For concreteness, we consider an admixture of open and closed highly excited strings in three noncompact dimensions with $N_D$ spacefilling D-branes.
We note that, as reviewed in appendix~\ref{sec:setup}, open (with closed) strings in 3 noncompact dimensions with spacefilling D-branes have a well defined canonical ensemble for any energy density with temperature asymptoting to the Hagedorn temperature at infinite density (systems of this nature are dubbed ``limiting'' in \cite{Abel:1999rq}).
The D-branes may also fill some of the compact dimensions.
Because the dense brane scenario described above has somewhat different thermodynamic properties and cosmological behavior, our focus here is on the BV and JJP box scenarios; we comment on the dense brane scenario in appendix \ref{sec:setup}.\\

As explained in~\cite{Frey:2023khe} (see also \cite{Lowe:1994nm,Lee:1997iz,Copeland:1998na}), an ensemble of highly excited strings in 3 noncompact and 6 small compact dimensions obeys Boltzmann equations
\begin{eqnarray}
\frac{\partial \tilde{n}_o(l)}{\partial t}&=&\, a_1N_D l\tilde{n}_c(l) - \frac{b_1}{2N_D}\frac{\tilde{n}_o(l)}{l^{3/2}}+\int_{l_c}^{l-l_c}dl'\left(\frac{b_2}{2N_D}\tilde{n}_o(l')\tilde{n}_o(l-l')
-a_2 N_D\tilde{n}_o(l)\right)\nonumber\\
&&+\int_{l+l_c}^\infty dl'\left(2a_2 N_D \tilde{n}_o(l')-\frac{b_2}{N_D}\tilde{n}_o(l)\tilde{n}_o(l'-l)\right)
+\cdots\nonumber\\
\frac{\partial \tilde{n}_c(l)}{\partial t}&=& \frac{b_1}{2N_D}\frac{\tilde{n}_o(l)}{l^{3/2}} -a_1N_D l\tilde{n}_c(l)+\cdots
\label{eq:boltzmann1}
\end{eqnarray}
to leading order in the string coupling. 
We describe the ensemble through the quantities $\Tilde{n}_{o,c}(l)$, the number of strings with lengths between $l$ and $l+dl$ per noncompact volume.\footnote{As opposed to the notation in~\cite{Frey:2023khe} which are total numbers and not densities.}
These terms describe open-open and open-closed endpoint interactions; closed-closed and other interactions in string interiors are higher order. The coefficients are
\begin{equation}\label{eq:int-coeffs}
a_1\simeq \frac{g_s M_s^2}{\Omega_\perp} ,\quad b_1\simeq \frac{g_s}{M_s^{1/2}\Omega_\|} ,\quad
a_2\simeq \frac{g_s M_s^2}{\Omega_\perp} ,\quad b_2 \simeq \frac{g_s}{M_s^2\Omega_\|} ,
\end{equation}
where $g_s$ is the string coupling and we have neglected phase space factors. The cutoff at short string length is $l_c\sim 9/M_s$~\cite{Manes:2001cs,Frey:2023khe}.
Note that the string length $l$ includes the extent of the string in the suppressed compact dimensions and is not the projected length in the noncompact directions. It therefore represents the total string energy.
In a scenario where the long strings are localized to a small region of the compact dimensions, $g_s$ and $M_s$ are the local string coupling and scale at that location. 
The factors $\Omega_\perp,\Omega_\|$ in coefficients $a_{1,2}$ and $b_{1,2}$ are respectively the compact volume perpendicular and along the branes that are filled by the long strings, as measured in string units. 
Up to factors of order unity, they are powers of $l_{KK}/\sqrt{\alpha'}$; if the SM is supported on D3-branes, $\Omega_\|\equiv 1$ for zero compact dimensions along the branes.

The equilibrium configuration reads
\begin{equation}\label{eq:equilibrium-condition}
\Tilde{n}_c(l)\simeq M_s^4\frac{e^{-l/L}}{(M_s l)^{5/2}}\, , \qquad \Tilde{n}_o(l)\simeq \frac{N_D^2 M_s^4\Omega_\|}{\Omega_\perp} e^{-l/L}\, ,
\end{equation}
where the expressions are valid up to order one factors\footnote{These are model dependent and, for open strings, can always be reabsorbed into the parameter $N_D$.} for $l\geq l_c$, and $L^{-1}=M_s^2(\beta-\beta_H)$ determines the temperature.
The energy densities are
\begin{equation}
\label{eq:enerden}
\rho \approx \rho_o = M_s^2\int_{l_c}^{\infty}{dl' \, l' \Tilde{n}_o(l')}\simeq 
\frac{(N_DLM_s)^2\Omega_\|}{\Omega_\perp} M_s^4 ,\quad
\rho_c \simeq M_s^{7/2}/l_c^{1/2}\simeq M_s^4\, .
\end{equation}
where here and in what follows we neglect $\mathcal{O}(l_c/L)$ corrections such as the ratio $M_sl_c\sim 9$. 
Open strings dominate the energy (and entropy) density when $N_DM_s L \gg \sqrt{\Omega_\perp/\Omega_\|}$.
In the JJP scenario, both volumes are $\Omega_\perp\sim\Omega_\|\sim 1$, so open strings are always dominant.
In the BV scenario with D$p$-branes, open strings dominate for $N_DM_s L \gg (M_sl_{KK})^{6-p}$; since the BV scenario assumes $L\gg M_sl_{KK}^2$, open strings always dominate for $p\geq 4$, but closed strings may dominate for a brief period at the end of the Hagedorn phase for D3-branes. 
In the rest of the main text, we therefore assume that open strings dominate and leave the cosmology and gravitational wave signature of dominant closed strings to future work. \\

Now consider a small perturbation $\delta\tilde{n}_c,\delta\tilde{n}_o$ to the equilibrium distributions (\ref{eq:equilibrium-condition}).
The rate at which the distributions return to equilibrium (equilibration rates) for strings in flat spacetime were given in \cite{Frey:2023khe}.
One important feature of equilibration is that longer strings return to equilibrium faster; that is, $\delta\tilde{n}_c(l),\delta\tilde{n}_o(l)$ vanish faster for larger $l$.
Also, because the endpoint interactions are lower order in string perturbation theory, open strings equilibrate among themselves and with closed strings much faster than closed strings equilibrate on their own (except for the nearly vanishing number of extremely long closed strings).
Therefore, closed strings can equilibrate more efficiently through their interaction with open strings.
Using the notation $\Gamma_{c,c}\, , \, \Gamma_{o,o}\, , \, \Gamma_{o,c}$ to indicate the equilibration rates for energy transfer in the closed and open and between the open-closed sectors for string length $l$, we find 
\begin{eqnarray}
\Gamma_{o,o}(l)&\simeq& \frac{g_s N_D M_s^2}{\Omega_\perp}\left(L+\frac{l}{2}\right)\gtrsim g_s N_D M_s^2 L/\Omega_\perp ,\nonumber\\ 
\Gamma_{o,c}(l)&\simeq& \frac{g_s N_D M_s^2 l}{\Omega_\perp} + \frac{g_s M_s}{2N_D \Omega_\| (M_s l)^{3/2}}
\gtrsim g_s N_D M_s^2 l/\Omega_\perp\sim g_s N_D M_s^2 L/\Omega_\perp\, ,
\label{eq:equilibration-rates}\\
\Gamma_{c,c}(l) &\simeq& \frac{g_s^2 l}{\Omega_\| \Omega_\perp}\left(\frac{\rho_c}{M_s^2}+\frac{M_s^2}{2(M_s l_c)^{1/2}} \right)\simeq 
\frac{g_s^2 M_s^2l}{\Omega_\|\Omega_\perp} \sim
\frac{g_s^2 M_s^2 L}{\Omega_\|\Omega_\perp} .
\nonumber
\end{eqnarray}
The last relation on each line is valid for typical strings in the ensemble, which are length $L>1/M_s$.
In addition, we neglect $\mathcal{O}(1)$ factors. 

\subsection{Hagedorn cosmology}\label{sec:cosmology}

In an expanding universe, dilution of energy density continually pulls a thermal system away from equilibrium; this is true for strings as much as for standard particle physics. 
Using standard arguments, the equilibrium configurations for strings are only valid in cosmology if the equilibration rates are much larger than the Hubble scale.
To make this more concrete, let us study the evolution equation for a fluctuation around the equilibrium distribution.
Following \cite{Copeland:1998na}, the straightforward generalization of the equations of~\cite{Frey:2023khe} reads generically 
\begin{equation}\label{eq:expandingboltzmann}
\frac{\partial n}{\partial t}+3H n + \frac{\partial (\dot{l} n)}{\partial l} = \textnormal{interactions} ,
\end{equation}
where $\dot l$ is the growth rate of the string's length with the expansion of the universe.\footnote{assuming that $\dot l=0$ at $l=l_c$.}
Together, the left-hand side of (\ref{eq:expandingboltzmann}) implies conservation of the total number of strings per comoving volume in the absence of interactions; the first and third terms make up the continuity equation in $l$-space.
We expect that strings at or longer than the Hubble scale will stretch with the expansion of the universe, while shorter strings may not, so $\dot l\sim Hl$. Therefore, as in particle physics, the linearized Boltzmann equations are a competition between expansion terms at scale $H$ and interaction terms with equilibration rate $\Gamma$ as described in the previous section. \\

We will assume that the string interaction rates (and therefore the equilibration rates) are the same in the cosmological spacetime as in flat spacetime. 
In particle physics, this assumption is justified because the interaction occurs over very small length and time scales; this is also true for splitting/joining or reconnection interactions at fixed points along the string(s).
However, the rates for a string to decay or for an open string to close also depend on the geometric probability for a random walk to self-intersect, which spacetime curvature could logically affect.
Nonetheless, at fixed proper time, the spatial curvature is extremely small in realistic cosmologies, so we expect that any effects should be negligible over the extent $\sqrt{L/M_s}$ of a typical string, and we believe our assumption to be reasonable. \\

The equilibration rates in (\ref{eq:equilibration-rates}) above are to be compared with the Hubble rate which, as the expansion of the Universe is sourced by the dominant contribution to the energy density (open strings), reads
\begin{equation}\label{eq:hubble}
H \simeq \frac{\sqrt{\rho}}{M_p}=N_D L M_s^2 \frac{M_s}{M_p}\sqrt{\frac{\Omega_\|}{\Omega_\perp}}\, ,
\end{equation}
where $M_p$ is the Planck scale of the effective 4D theory.
An important consistency condition to neglect string theoretic corrections to the expansion of the Universe is that $M_s/H \gg 1$, and thus our setup is only consistent provided
\begin{equation}\label{eq:hubble2}
    N_D L M_s \, \frac{M_s}{M_p} \sqrt{\frac{\Omega_\|}{\Omega_\perp}}\ll 1 \, .
\end{equation}
Since we are assuming $N_D L M_s\sqrt{\Omega_\|/\Omega_\perp}\gg 1$ to impose open string dominance, we need to impose a hierarchy between $M_s$ and the 4-dimensional Planck scale.
This is possible in JJP models with large extra dimensions (and weak warping), which have concrete realizations in string theory~\cite{Balasubramanian:2005zx}, or compactifications that feature highly warped regions~\cite{Giddings:2001yu,Giddings:2005ff}, provided the relevant physics occurs in such regions.
We will assume this in the following, noting that these setups are the most frequent scenarios for model building, though we also check below that BV scenarios can satisfy even stricter consistency conditions. 
Note that (\ref{eq:hubble2}) also requires that the energy of a typical long string is $M_s^2L\ll M_p$ for strings in the JJP box. \\

The condition $H\ll M_s$ for consistency of the effective theory is robust, but it is not the most stringent. For Kaluza-Klein scale $M_{KK}$ (or the scale of some other tower of states as discussed in \cite{Ooguri:2006in}), there is the model-dependent condition $H\ll M_{KK}$, or 
\begin{equation}\label{eq:hubble3}
N_D L M_s \frac{M_s}{M_p}\sqrt{\frac{\Omega_\|}{\Omega_\perp}}\ll \frac{M_{KK}}{M_s} .
\end{equation}
For models of large extra dimensions of linear scale $l_{KK}$ (without a high degree of anisotropy or inhomogeneity), $M_{KK}\sim 1/l_{KK}$ and $M_p\sim M_s^4 l_{KK}^3/g_s$. 
In BV models with D$p$-branes, we require $N_D M_s L\ll (M_sl_{KK})^{8-p}/g_s$, which is compatible with the requirement that $L_{rms}\gg l_{KK}$ for $p<6$ (and with the open string dominance condition).
For large volume JJP models, the JJP box has effective compact volumes of order 1, which also leads to a loose bound $N_D M_s L\ll (M_sl_{KK})^{2}/g_s$.
On the other hand, the expectation for strongly warped compactifications with the long strings in the warped throat is for $M_{KK}\lesssim M_s$, so (\ref{eq:hubble2}) is only slightly modified. 

If the Hagedorn phase follows inflation at scale $H_{inf}$, consistency of the effective theory during inflation also provides a somewhat stronger condition, as described in the introduction (though note that the string scale $M_s$ may be larger during inflation due to shifted moduli expectation values \cite{Frey:2005jk}).\\

It follows that the equilibration rates involving highly excited strings have the behaviour
\begin{eqnarray}
\frac{\Gamma_{o,o}}{H}&\simeq& \frac{g_sM_p}{M_s\sqrt{\Omega_\| \Omega_\perp}}\equiv R\geq 1 \, , \quad \frac{\Gamma_{o,c}}{H}\simeq \frac{l}{L(t)} \frac{g_sM_p}{M_s\sqrt{\Omega_\| \Omega_\perp}}=R\frac{l}{L(t)}\, ,\nonumber\\
\frac{\Gamma_{c,c}}{H}&\simeq& \frac{g_s^2 l}{N_D L(t)\Omega_\|^{3/2}\Omega_\perp^{1/2}}\frac{M_p}{M_s} =\left(\frac{g_s l}{N_D L(t)\Omega_\|}\right)R\, .
\label{eq:rate-hubble-ratio}\end{eqnarray}
Here, $L(t)$ is the time dependent length scale of the string gas, which we recall determines the temperature.
In terms of the parameters of the compactification, $R=e^{-A_0}(V_w/\alpha'^3)^{1/2}(\Omega_\| \Omega_\perp)^{-1/2}$, where $V_w$ is the (possibly warped) volume of the entire compactification, and we are allowing for a nonzero warp factor $A_0$, which will be generically present in realistic compactifications, albeit not necessarily with an extreme value (recall that the Jackson-Jones-Polchinski argument holds even for weak warping).
In many situations of phenomenological interest, the ratio $R$ should be large, so the long open strings stay in equilibrium at all times until the temperature drops sufficiently below the Hagedorn temperature that $L(t)\lesssim l_c$ (when the strings are no longer highly excited).
However, in the BV scenario, we ignore warping, and the long strings spread through the entire compactification, so the volume factors cancel and $R\sim 1$. 
The strings are therefore marginally in equilibrium, so the thermodynamics is sensitive to order 1 phase space factors in the equilibration rates, and the full cosmological Boltzmann equations are likely necessary. We therefore leave a more detailed study of the BV scenario to future work.\\

The ratios of equilibration rates with respect to the Hubble scale therefore reveal a coherent setup in which open string degrees of freedom always thermalise.
Long closed strings ($l>l_c\sim 1/M_s$) also equilibrate through their interaction with open strings because we have also assumed $M_s^2 L(t)\ll M_p$ for consistency of the EFT\footnote{In absence of branes, whether only closed strings can reach thermal equilibrium is a model dependent question, but in this scenario the canonical ensemble breaks down at order one energy densities and the description is less clear.}.
It is worth noting that equilibration also occurs faster than density perturbations grow; the Jeans instability in flat space occurs over a parametrically identical time scale as the Hubble expansion, and density perturbations grow more slowly (as power laws) in cosmology.
After sufficient dilution, only massless open string fields (which we assume to be the SM or an extension thereof, as is common in string model building) remain in equilibrium, providing a natural window into the hot big bang.\\

However, in section~\ref{sec:gw-strength} we will see that the above results do not apply for massless closed strings, which we will argue do not equilibrate.
Heuristically, long strings are confined to a box of size $\Omega_\perp$, so dimensional reduction of their interaction coefficients is carried in this box (c.f. eq.~\eqref{eq:int-coeffs}).
Gravitons and other massless closed strings are however not confined to the JJP box, as we will explain in appendix~\ref{sec:setup}. Since the supergravity equations imply the graviton wavefunction spreads through the whole compact space, dimensional reduction results in Planckian suppression.\\

Before concluding this section, let us study conservation of entropy.
The exponentially large number of degrees of freedom that release their entropy into the plasma as the Universe expands implies that the temperature takes a long time to drop: it is the inverse temperature difference $L$ that scales with the scale factor (as opposed to the more standard $T\sim a$ in weakly coupled particle thermodynamics).
To see this, notice~\cite{Manes:2001cs,Frey:2023khe} that the typical strings in the ensemble are very nonrelativistic, so that the pressure $P\sim \sqrt{\rho}M_s^2$ is negligible, and the entropy density of the system is thus $s=\beta \rho = \rho \lr{\beta_H+1/(M_s^2L)}\simeq \beta_H \rho $ \cite{Abel:1999rq}.
Therefore, conservation of comoving entropy $s a^3=\text{const.}$ requires $\rho \sim a^{-3}$, as appropriate for the energy density of a nonrelativistic gas.
The Hagedorn phase thus behaves cosmologically like a period of \textit{early matter domination}.
This will be very important when we compare the GW spectrum arising in the Hagedorn phase with that of the Standard Model in standard cosmology.
We are therefore able to track the inverse temperature difference
\begin{equation}\label{eq:l-evolution}
L(t)=L_*\lr{\frac{a_*}{a(t)}}^{3/2}\, ,
\end{equation}
where the expression is valid whenever $L\gg 1$.

\section{Gravitational wave emission from long strings}\label{sec:gw-emission}

The goal of the present article is to compute the GW spectrum arising from the Hagedorn phase.
To do so, we will study the decay rate of a typical string at a given length by emission of gravitons (massless closed strings), averaging over the initial states at the given mass level.
A key point is that the effects of realistic compactifications are different for highly excited strings and gravitons. This affects the overall scaling of the graviton emission amplitude as compared to the long string interactions of the previous section, such as the splitting of a long open string into two other long open strings. 
Heuristically, the probability of a highly excited string decaying into another highly excited string and a graviton is proportional to the square of the disk amplitude, which reads
\begin{equation}
    \mathcal{A}=\langle \mathcal{V}_{\tti{HES}} \mathcal{V}_{\tti{HES}} \mathcal{V}_g \rangle ,
\end{equation}
where $\mathcal{V}_{\tti{HES}}$ and $\mathcal{V}_g$ are the vertex operators for highly excited strings and gravitons respectively. 
In a product compactification (such as the first scenario discussed in \ref{sec:setup}),the vertex operators are $\mathcal{V}\sim g/\sqrt{V} e^{ik x}\mathcal{O}$, with $\mathcal{O}$ a product of derivatives of $X$, $g$ the open or closed string coupling as appropriate, and $V$ the compact volume.
After accounting for the path integral normalization (including zero modes), $\mathcal{A}\sim g_s/\sqrt V$.
However, suppose the long strings are localized to a string-scale volume $V_{\tti{HES}}$. 
Then the graviton-emission probability is suppressed by a factor $V_{\tti{HES}}/V$ in comparison.
Furthermore, in warping, the graviton vertex operator should have a nontrivial profile in the compact dimensions, which we expect to suppress the amplitude by the warp factor.\\

In the absence of a worldsheet description for compactifications with moduli stabilization, we will take another approach to make the above arguments rigorous.
We begin in section~\ref{sec:gw-strength}, where we study graviton emission from a nonrelativistic object with tension $T_p$ that is localized in the compact dimensions of a warped background (hence at zeroth order in a gradient expansion). 
For long strings extended in the noncompact dimensions, we find that the interaction rate is suppressed by a factor $M_s/M_p$, where $M_s$ is the warped string scale and $M_p$ is the Planck scale as above.
The remainder of the emission amplitude is the matrix element of the graviton vertex operator in the initial and final states of the highly excited string in the flat noncompact dimensions (to the extent that we can ignore cosmological expansion). \\

Having understood the relative strength of the interactions, in section~\ref{sec:GW-spectrum} we review the computation of the decay rate of a typical string into gravitons, which has been carried out in the bosonic~\cite{Amati:1999fv} and supersymmetric~\cite{Kawamoto:2013fza} case at leading order in string perturbation theory in flat backgrounds.
Recently, \cite{Firrotta:2024fvi}, using arguments from the optical theorem, has raised a technical subtlety regarding the analogous computation for the emission of massless open strings by open strings and the emission of massless closed strings by closed strings (which therefore does not include our case of interest).
We will not have anything new to say about this disagreement and will restrict ourselves to parametrising the resulting GW spectrum in a sufficiently general way that can accommodate the main lessons learned from both points of view (we leave a detailed
resolution of the discrepancy to future works).
We will see that our conclusions are robust enough to accommodate this disagreement and further corrections. 
The general result is that the emission spectrum features a greybody spectrum at the Hagedorn temperature, $T_H \sim M_s$, suppressed by $(M_s/M_p)^2$.
Using the leading order factorization of the worldsheet CFT, we find that \textit{the dependence of the emission rate on the details of the compactification is cancelled out by the averaging procedure}, rendering our results robust against model dependence.\\

In section~\ref{sec:grav-compact} we merge the results from both sections and apply them to our case of interest: (four-dimensional) graviton emission from highly excited open strings in the presence (or not) of warping.
We will find that the computations carry through by using the dimensionally reduced coupling and the warped string scale.
We also carry out the computation of the decay rate.
\\

\subsection{Quantum mechanical analysis}\label{sec:gw-strength}

As argued above, the emission of gravitons by highly excited strings should be suppressed due to the localization of the long strings. 
In fact, as expected for a gravitational process, the emission rate will be Planck suppressed. 
Since there is not yet a full worldsheet description of compactifications with warping (plus flux and other ingredients needed for moduli stabilization), we will determine the effects of warping using a quantum mechanical analysis, treating the initial and final long strings as a excited states of a single nonrelativistic object.\\

The spirit of the analysis is the same as that used to determine the interaction of light with matter (see e.g. \cite{Weinberg_2015}). The starting point 
is the action of a single long, nonrelativistic fundamental string coupled to 10D gravity. We linearise gravity; the Hilbert space of the entire system is a tensor product of the states of the fundamental string and gravitons.   Quantisation
of the gravitational sector can be carried out by canonical
methods, and the normalisation involved in this
process sets the strength of gravitational interactions
of the fundamental string. Furthermore, the matrix element that determines
the S-matrix for graviton emission factorises; it is the product of a matrix element in the gravitational sector 
(which is a free field theory matrix element) and a matrix element of the quantum fundamental string.\\ 

Let us begin by considering the action of an object with $p$ spatial dimensions that is minimally coupled to gravity (in a gauge where worldvolume time is the same as the overall time coordinate):
\begin{equation}
S=\int dt \lr{ \frac{M_{10}^{8}}{2} \int d^3x\,d^6y\,\sqrt{-g} \left[ R_{10}+\cdots\vphantom{\frac 12}\right] +T_p\int{d^p \sigma{\sqrt{-\gamma}}}}\, ,
\end{equation}
where $M_{10}$ is the $10$-dimensional Planck mass, $T_p$ is the tension of the localized object, and $\gamma_{ab}$ is the pullback of the spacetime metric $g_{MN}$ on the worldvolume of the object.
The $\cdots$ represent the contribution of additional fields to the bulk action, which are responsible for moduli stabilization and the appearance of a warp factor. 
Because we assume moduli are stabilized, the string and Einstein frames are identical for our purposes.\\

The background spacetime metric of interest takes the form
\begin{equation}\label{eq:metric1}
ds^{2} = e^{2A(y)} \eta_{\mu \nu} dx^{\mu} dx^{\nu}
+ e^{-2A(y)} \tilde{g}_{mn} dy^{m} dy^{n}
\end{equation}
Furthermore, any metric with $\eta_{\mu \nu} \to g_{\mu \nu} (x)$, where $g_{\mu \nu}$ is Ricci flat, continues to satisfy the equations of motion \cite{Giddings:2001yu}. Given this, the fluctuations associated with the
four dimensional graviton correspond to taking  $ \eta_{\mu \nu} \to \eta_{\mu \nu} + h_{\mu \nu}(x)$, with $h_{\mu \nu}(x)$ solving the four dimensional Lichnerowicz equation.
 Dimensional reduction to four dimensions with a metric ansatz of the form \eqref{eq:metric1} with
$\eta_{\mu\nu}\to g_{\mu\nu}(x)$
chosen to capture the degrees of freedom associated with the 4d graviton yields
an effective action
\begin{equation}
\label{gravee}
      S  \supset \int{dt \lr{\frac{M_{p}^{2}}{2}\int{d^3x\,\sqrt{-g(x)} R (x)}+
T_p\int{d^p \sigma{\sqrt{-\gamma}}}}}\,
\end{equation}
where $M_{p}$ is the four dimensional Planck mass. This is related to the ten-dimensional Planck mass by a
factor of the warped volume
$$
M^2_{p} = M^8_{10} \int d^{6}y  \sqrt{\tilde{g}}e^{-4A}.
$$
We note that the contribution of highly warped regions
to the warped volume is typically negligible, so its value
is set by the overall volume of the  compactification
and is insensitive to the minimum value of the warp factor
(see e.g. \cite{Giddings:2005ff}).\\

Now suppose that the object has a small spread in the compact dimensions around a local minimum of the warp factor $y =y_0$.
The form of the effective action  \eqref{gravee} has two important implications.
First, the tension of the object is warped down $T_{p} \to e^{(p+1)A(y_0)} T_{p}$, and this warped tension governs the gravitational dynamics of the localised object (at leading order in a derivative expansion for the warp factor). 
The case of interest is a localised fundamental string ($p=1$), which therefore has an effective string scale
$M_s\equiv e^{A(y_0)}/\sqrt{\alpha'}$ as measured with respect to the time coordinate $t$. This is the same warped string scale that we have used previously.\\

One can also obtain the coupling of a localised
string to gravitons (generalizing to higher-dimensional objects is trivial). Taking a perturbed metric $g_{\mu \nu} (x) =  \eta_{\mu \nu} + h_{\mu \nu} (x)$,
the leading form of the action \eqref{gravee} is
\begin{equation}
S=\int{dt\,\lr{ \frac{M_{p}^2}{2}
\int{d^3x\, \lr{\partial h(x)}^2}+
\frac{M_s^2}{2\pi}\int{d\sigma{\sqrt{-\hat{\gamma}_{0}}}}+
\frac{M_s^2}{2\pi}\int {d\sigma \sqrt{{- \hat{\gamma}_{0}}} \,h_{\mu \nu}(x(\sigma))x^\mu_a x^\nu_b \hat{\gamma}_{0}^{ab}}}}\, ,
\end{equation}
where $(\partial h)^2$ schematically indicates the kinetic term for the graviton field with indices appropriately contracted, $\hat{\gamma}_{0}$ is the pullback metric obtained from the Minkowski metric $(\eta_{\mu \nu})$, and $x_a^\mu\equiv \partial x^\mu/\partial\sigma^a$ for $(\sigma^0,\sigma^1)=(t,\sigma)$. 
The first term on the string worldsheet determines the states of the long string, while the second is the leading interaction with 4D gravity.
Note that the canonically normalised graviton field $\hat{h}_{\mu\nu}$ is obtained by the rescaling $\hat{h}_{\mu \nu}=h_{\mu \nu} M_{p}$. 
We see that the interaction between fluctuations of the localised object and
the 4D graviton is indeed suppressed by the 4D Planck scale.\\

Thus the system consists of two sectors (gravitons and the fundamental string) coupled by a linear interaction. 
This is exactly the same situation that one encounters while discussing the interactions of light with matter. 
We will follow the treatment of \cite{Weinberg_2015} to analyse graviton emission when the fundamental string transitions from state $|A \rangle$ to $|B \rangle$ (for the composite system, the final state is $|B,g (\vec{k}, e_{\mu \nu})\rangle$, $\vec{k}, e_{\mu \nu}$ being the momentum and polarisation of the outgoing graviton). 
The S-matrix for the process can be computed using the interaction picture and is given by the matrix element
$$
{\mathcal{M}}_{A\to B,g}  = \frac{M_s^2}{2\pi M_{p}} \int d^2\sigma \left\langle B,g(\vec{k}, e_{\mu \nu}) \left| \hat{h}_{\mu \nu}(x(\sigma))
\sqrt{-\hat{\gamma}_{(0)}} x^\mu_a x^\nu_b \hat{\gamma}_{0}^{ab}\right|A\right\rangle .
$$
The Hilbert space factorises ($|B,g\rangle=|B\rangle \otimes |g\rangle$), so the gravity sector matrix element is easily evaluated by expanding the graviton in terms of creation and annihilation operators. This leads to 
\begin{equation}
\label{finalmat}
{\mathcal{M}}_{A\to B,g}  = \frac{M_s^2}{2\pi M_{p}} e_{\mu \nu} \int d^2\sigma \left\langle B \left| e^{ik\cdot x(\sigma)}
\sqrt{-\hat{\gamma}_{(0)}} x^\mu_a x^\nu_b \hat{\gamma}_{0}^{ab}\right|A\right\rangle .
\end{equation}
In terms of the quantum mechanics of the long string, this is the transition amplitude as given by time-dependent perturbation theory. \\

However, the attentive reader might notice the similarity between the operator in the matrix element in \eqref{finalmat} and the form of the graviton vertex operator. 
Indeed, up to normalization, both are the same after expressing $\gamma$ in conformal gauge and $\sigma$ in holomorphic coordinates.
Therefore, we can use the (unnormalized\footnote{ In toroidal compactifications at weak coupling and large volume the
normalisation of vertex operators (as described in the beginning of this section) can be obtained explicitly, yielding $\mathcal{M}\sim g_s/\sqrt{V}\sim M_s/M_p$ in agrement with our general argument.}) flat spacetime string perturbation theory amplitude with appropriate polarizations in \eqref{finalmat} to determine the emission rate.
We will later argue that corrections due to cosmological expansion can be neglected.
In all, we see that the graviton emission is that of a fundamental string whose tension is given by the warped string scale, and the rate is suppressed by the Planck scale.

\subsection{Graviton emission amplitude in flat backgrounds}\label{sec:GW-spectrum}

We thus conclude that we can use the results of~\cite{Amati:1999fv,Kawamoto:2013fza} in a warped background (under the usual assumptions), provided we modify the couplings and the effective string scale appropriately.
Warped backgrounds also generically include NSNS and RR flux, but we note that the effects of flux in worldsheet computations are suppressed by higher powers of $g_s$~\cite{Cho:2023mhw}, so we can ignore them.\\

Let us thus review the computations in~\cite{Amati:1999fv,Kawamoto:2013fza}.\footnote{We note again that a recent complementary calculation using the optical theorem by \cite{Firrotta:2024fvi} differs from these results. We will see in section \ref{sec:sto} that the disagreements in the literature do not modify our conclusions substantially. We leave a detailed resolution of this discrepancy
for future works.}
The strategy is to obtain the probability of a highly excited string at level $N$ to decay into a graviton with frequency $\omega$ and another highly excited string at level $N'$.
Summing over final states $\lbrace\Phi_{N',i}\rbrace$ at level $N'$ and polarizations $\xi$, and averaging over initial states $\lbrace{\Phi_{N,j}}\rbrace$ at level $N$ renders the \textit{averaged, semi-inclusive square amplitude} $F(\omega)$
\begin{equation}
    F(\omega)=\frac{1}{\mathcal{G}(N)}
    \sum_{i,j,\xi}
    |\langle \Phi_{N',i}|V_\xi (k)^*|\Phi_{N,j} \rangle |^2
    =\frac{1}{\mathcal{G}(N)}\sum_{i,j,\xi}
    \langle \Phi_{N',i}|V_{\xi}^* (k)|\Phi_{N,j} \rangle
    \langle \Phi_{N,j}|V_{\xi} (k)|\Phi_{N',i} \rangle \, ,
\end{equation}
where states and operators are to be understood as not carrying zero-modes as usual and $k$ is the wavevector of the graviton with frequency $\omega$.
To compute the amplitude, we trade the sums by a trace over the Fock space of the oscillator modes by inserting projectors at level $N$:
\begin{equation}
    \hat{P}_N \equiv \oint \frac{dz}{2\pi i z}z^{\hat{N}-N} \, , \qquad \sum_{j} | \Phi_{N,j} \rangle = \sum_{\Tilde{N},j} 
    \hat{P}_N|\Phi_{\Tilde{N}\, ,j} \rangle \, ,
\end{equation}
and thus the computation reduces to
\begin{equation}
    F=\frac{1}{\mathcal{G}(N)}\sum_\xi \oint{ \frac{d z}{2\pi i z} z^{-N}} \oint{\frac{dz'}{2\pi i z'}z'^{-N'}} \text{Tr} [V^\dagger _\xi (k,1) \, z'^{\hat{N}} V_\xi (k,1) z^{\hat{N}} ] \, .
\end{equation}
The resulting trace can be computed by textbook methods, provided the graviton vertex operator is identified appropriately.
On that matter, note that we are considering the graviton production from an \textit{open} superstring, and thus to perform the computation we write
\begin{equation}
    V_\xi (k,e^{i\tau}) = \int_0^\pi {\frac{d \sigma}{\pi} :V_{L,\xi}(k_L,e^{i(\tau+\sigma)})::V_{R,\bar{\xi}}(k_R,e^{i(\tau-\sigma)}):}\, ,
\end{equation}
where we have decomposed the polarization of the graviton into $\xi \bar{\xi}$, and a zero mode substraction should be understood.
The full computation can be found in~\cite{Kawamoto:2013fza}.
Here we simply quote the part of the vertex operator that yields the dominant contribution:
\begin{equation}
    V_{B,\xi}(k,z)= \xi^i \dot{X}^i(z) e^{i k\cdot X(z)} + \text{others}\, .
\end{equation}
The computation thus involves (noting $z'^{\hat{N}} V(k,1)z'^{-\hat{N}}=V(k,z')$) computing the following quantity:
\begin{gather}
    F=\frac{1}{\mathcal{G}(N)}\sum_\xi \oint{ \frac{d w}{2\pi i w} w^{-N}} \oint{\frac{dv}{2\pi i v}v^{N-N'}}
    \int_0^\pi {\frac{d\sigma}{\pi}}\int_0^\rho{\frac{d\rho}{\pi}} \, 
    \text{Tr} [\hat{O} w^{\hat{N}} ] \\
    \hat{O}\equiv e^{ \bar{\xi}^* \cdot \dot{X}(e^{-i\rho})-i k \cdot X(e^{-i\rho})}
    e^{\xi^* \cdot \dot{X}(e^{i\rho})-i k \cdot X(e^{i\rho})}\,
    e^{\xi \cdot \dot{X} (ve^{i\sigma})+ i k \cdot X(ve^{i\sigma})}
    e^{\bar{\xi}\cdot \dot{X}(ve^{-i\sigma})+ik \cdot X(v e^{-i\sigma })}\, ,
\end{gather}
where $w=zz'$ and $v=z'$.
Eventually, the result is of the form 
\begin{equation}
    \frac{1}{\mathcal{G}(N)}\oint \frac{dw}{2\pi i w}w^{-N'} f(w, N-N') Z(w)\, ,
\end{equation}
with $Z(w)$ the partition function of the theory, and for the present computation~\cite{Kawamoto:2013fza} 
\begin{equation}
    f(w,N-N') \simeq \frac{(N-N')^2}{\lr{1-w^{(N-N')/2}}^2}\, ,
\end{equation}
where we have neglected terms of order $\mathcal{O}(\omega \sqrt{\alpha'}/\sqrt{N})$ and higher.
Let us, however, keep $f(w,N-N')$ general except for the assumption that it does not dramatically affect the well-known saddle of $Z(w)$ at $\log (w) \to -M_s/(2\sqrt{N'}T_H)$.
If so, the remaining integral can be performed in a similar way to the computation of the number of states at level $N$ (times the function evaluated at the saddle), yielding
\begin{equation}\label{eq:prob-ratio-G}
    F=\frac{\mathcal{G}(N')}{\mathcal{G}(N)} f(e^{-M_s/(2\sqrt{N'}T_H)},N-N')\, .
\end{equation}
Let us now stop to comment on the generality of this result.
If the worldsheet CFT factorises and the vertex operator of the graviton only acts on the free CFT, we observe that the trace breaks into 
\begin{equation}
    \text{Tr}[\hat{O}w^{\hat{N}}]=
    \text{Tr}_{\tti{free}}[\hat{O}w^{\hat{N}_f}]
    \text{Tr}_{\tti{int}}[w^{\hat{N}_c}]=
    \text{Tr}_{\tti{free}}[\hat{O}w^{\hat{N}_f}] Z_{\tti{int}}(w)\, ,
\end{equation}
where we have divided the number operator $\hat{N}=\hat{N}_f+\hat{N}_c$ into the free (noncompact) and compact part, and analogously for the trace.
It follows that the compact part simply contributes with its partition function and the known factor $f(w,N-N')Z_{\tti{free}}(w)$ arises from the free contribution, which we can compute.
In addition, because $Z(w)=Z_{\tti{free}}(w) \, Z_{\tti{int}}(w)$ grows exponentially, the saddle point approximation can always be performed, and the details of the compactification only appear in $F$ through $\mathcal{G}(N)$.
The result in Eq.~\eqref{eq:prob-ratio-G} is therefore true in any background allowing for a worldsheet CFT that factorises into compact and noncompact parts. 
While we expect that the CFT of a warped compactification does not factorize, this amplitude will factorize to leading order in a gradient expansion of the warp factor because the warp factor is constant at the (localized) long string position in that approximation.\\

Since the emission rate requires conservation of energy, we may approximate $N-N' \sim 2\sqrt{ N} \omega/M_s$, $\sqrt{N}-\sqrt{N'}\simeq \omega/M_s$.
Because in general $\mathcal{G}(N)=b\, N^a e^{\beta_H M_s \sqrt{N}}$, with $a$ and $b$ constants, the polynomial parts of $\mathcal{G}(N)$ and $\mathcal{G}(N')$ are equal at leading order in $\omega/(M_s\sqrt{N})$, and so the leading order result is
\begin{equation}\label{eq:amati-russo}
     F\simeq  4(l\, \omega)^2 \frac{e^{-\omega/T_H}}{(1-e^{-\omega/(2T_H)})^2}\, ,
\end{equation}
where $l=M/M_s^2 \simeq \sqrt{N}/M_s$.
The emission of gravitons is thus predicted to follow a greybody spectrum peaking close to the Hagedorn temperature.

\subsection{Graviton radiation in compactifications}\label{sec:grav-compact}

Let us now put together the above results to find the decay rate of long strings by 4D graviton emission in realistic setups.
We argued in Sec.~\ref{sec:gw-strength} that the emission amplitude in realistic compactifications is the graviton emission amplitude in flat spacetime up to normalization, including $M_s/M_p$ suppression.
We reviewed the computation of this flat space amplitude in Sec.~\ref{sec:GW-spectrum} to leading order in string perturbation theory.
We are thus in a position to compute the decay rate.\\

The square amplitude $F$ gives the decay rate for a long string of level $N$ (and mass $M^2=M_s^2(N-1)$) to a string of level $N'$ ($M'^2=M_s^2(N'-1)$) and a 4D graviton with frequency $\omega$, averaged over the states of level $N$ and summed over all states of level $N'$.
Using the normalisation of section \ref{sec:gw-strength}, the emission rate for gravitons with frequency from $\omega$ to $\omega+d\omega$ for fixed $N,N'$ is
\begin{equation}\label{eq:rate-Nprime}
d\Gamma_{N\to \omega N'} = 
\frac{\omega}{32\pi^3 M}\frac{M_s^4}{M_p^2} F
\frac{\delta(M-\sqrt{M'^2+\omega^2}-\omega)}{\sqrt{M'^2+\omega^2}} d\omega .
\end{equation}
However, we would like the total emission rate of gravitons within a given frequency bin from a string of level $N$, so we should sum over the product state levels $N'$. 
At large $N'$, we can convert the sum to an integral over the mass using $dN'=2M' dM'/M_s^2$, so 
\begin{equation}\label{eq:rate-per-freq}
\frac{d\Gamma}{d\omega} = \frac{1}{16\pi^3}\left(\frac{M_s}{M_p}\right)^2 \frac{\omega}{M}\int dM' \frac{M'F}{\sqrt{M'^2+\omega^2}}
\delta(M-\sqrt{M'^2+\omega^2}-\omega)=
\frac{1}{16\pi^3}\left(\frac{M_s}{M_p}\right)^2
\frac{\omega F(l,\omega)}{M} ,
\end{equation}
where we now use energy conservation to write $F$ as a function of the initial string mass and outgoing graviton frequency.\\

Using the square amplitude \eqref{eq:amati-russo},
\begin{equation}\label{eq:rate}
\frac{d \Gamma_{o,g}}{d\omega}= \Tilde{A} \lr{\frac{M_s}{M_p}}^2 lM_s (\omega/T_H)^{3} \frac{ e^{-\omega/T_H}}{\lr{1-e^{-\omega/2T_H}}^2}\, ,
\end{equation}
with $\Tilde{A}=(T_H/M_s)^3 /(4\pi^3)$.
It follows that highly excited typical strings radiate massless strings like greybodies at the Hagedorn temperature, with a strength that depends on the ratio between the local string scale and the Planck scale.
This is the first hint of an exciting result: measuring the gravitational radiation arising from a Hagedorn phase would provide information about the local string scale.\\

A possible concern is that the cosmological expansion could modify the emission rate \eqref{eq:rate}, at least for strings longer than the Hubble radius. 
This is not usually an issue in particle physics because particle interactions are local processes. 
However, an interpretation of the single power of $l$ in \eqref{eq:rate} is that the long string has a constant graviton emission rate per unit length. 
In other words, each local segment emits gravitons independently of the rest of the string. 
As a result, we can think of graviton emission, like any standard particle physics process, as occurring locally, so it should be unaffected by the expansion of the universe.
In addition, corrections to the emission rate should be small for strings well contained within a Hubble radius. This is true for typical strings in the gas when $HL_{rms}\ll 1$, which is equivalent to $M_s^2L \ll M_p (M_s/M_p)^{1/3}$, a somewhat more stringent requirement than consistency of the EFT. Note though that it is likely to be satisfied at the end of the Hagedorn phase, which we will see is the most important era for the gravitational wave spectrum.
Nonetheless, we will later parameterise the emission spectrum in a way that could account for possible corrections or other dependence on $l$.\\

It would be very interesting to study the production of more model-dependent (but still generic) species in this setup, like gravitini, axions or closed string moduli, which are accessible by similar calculations.
In particular, the geometric moduli (like the volume modulus) are just polarizations of the 10D graviton in the compact dimensions.
In addition, the decay rates of long strings by emission of other massless strings (not only in the NSNS sector) are known in flat backgrounds.
According to~\cite{Kawamoto:2013fza}, they all behave like greybodies (with different greybody factors) at the Hagedorn temperature.
At this order in perturbation theory, closed string production from a given string is always proportional to the length of the mother string, and open string production is length-independent.
(For comparison, note that the decay rates quoted in~\cite{Kawamoto:2013fza} specify the level $N'$ of the product string, as in our \eqref{eq:rate-Nprime}.)\\

As a last remark, this computation allows us to show that gravitons never equilibrate.
We want to compare the total graviton production rate with the Hubble scale.
Integrating against $\omega$ gives an order one number, so that $\Gamma \sim (M_s/M_P)^2 l$.
We also need to integrate over all possible source strings for the gravitons, and this renders (due to the $l$ factor) a total production rate proportional to the energy density of the bath,
\begin{equation}
\Gamma_{g} \sim \lr{\frac{M_s}{M_p}}^2 \frac{\rho_b}{M_s^4} M_s \, .
\end{equation}
The Hubble scale is given by $H\sim \sqrt{\rho_b/M_s^4} M_s^2/M_p$ .
In a background with energy density $\rho \sim (N_D M_s L)^2 M_s^4$, as expected for long open strings, we find
\begin{equation}
\frac{\Gamma_g}{H}\sim (N_D M_s L)\lr{\frac{M_s}{M_p}}\, .
\end{equation} 
This is the same parametric behaviour as $H/M_s$, which needs to be much smaller than one in order for $\alpha'$ corrections to be negligible.
We thus conclude that gravitons never reach equilibrium in this scenario, and massless closed strings are produced as an out-of-equilibrium process.
In the next section we compute the GW spectrum arising from such out-of-equilibrium processes, and compare it to the analogous process in field theory.

\section{The spectrum of the stochastic gravitational wave background}
\label{sec:sto}

  In this section, we will compute the spectrum of the stochastic gravitational wave background produced from the decay of long open
  strings in a Hagedorn phase\footnote{During the Hagedorn epoch, the energy density in open strings is much larger than that in closed strings, see Eq. \eqref{eq:enerden}. Hence, the dominant channel for graviton production is the one from decay of open strings. } . We will use our results on the cosmology of the Hagedorn phase (section \ref{sec:thcosmo}) and the 
  decay rates of long open strings (section \ref{sec:gw-emission}) as key inputs.

 Let us begin by setting up some basic notation and conventions. Cosmological backgrounds of gravitational waves are usually expressed in terms of their fractional energy density per logarithmic frequency intervals, $h^2 \Omega_{\tti{GW}}(f_0)$~\cite{Caprini:2018mtu,Aggarwal:2020olq}:
\begin{equation}
\rho_{\tti{GW}}^{(0)} \equiv \int_0^\infty{d \log f_0 \, \rho_g^{(0)}(f_0)}  , \qquad 
h^2\Omega_{\tti{GW}} (f_0) \equiv h^2\frac{\rho_g^{(0)}(f_0)}{\rho_c}=\frac{15}{\pi^2}h^2\Omega_\gamma \frac{\rho_g^{(0)}(f_0)}{T_0^4}\, 
\label{phenoGW}
\end{equation}
where $\rho_{\tti{GW}}^{(0)}$ is the total energy density in gravitational waves.  In the last equality we have used the fractional energy density in photons $\Omega_\gamma=\pi^2 T_0^4/15\rho_c=2.47\cdot 10^{-5}/h^2$~\cite{Planck:2018nkj} to express the critical density $(\rho_c)$ in terms of the temperature of the CMB, $T_0$.
We will use angular frequencies ($\omega=2\pi f$) in section~\ref{sec:total-spectrum} for computational simplicity, but will use frequency today $f_0$ instead when comparing with observations in section~\ref{sec:features}.

\subsection{Total gravitational wave spectrum}
\label{sec:total-spectrum}

Next, let us turn to  computing the  spectrum
of  gravitational waves. Since gravitons never reach thermal equilibrium and propagate freely after their production, it is straightforward to write an evolution equation for the total energy density in gravitational waves $\left(\rho_{GW} (t)\right)$:
\begin{equation}
\label{eq:gwev}
\frac{\partial \rho_{GW}}{\partial t}+4H\rho_{GW}= \int_{l_c}^{\infty} \int_0^\infty{ \omega \frac{d \Gamma_{o,g}}{d \omega} \Tilde{n}_o(l) \, d\omega \, dl} \, .
\end{equation}
The evolution equation has two elements: firstly, the redshifting of the energy density in gravitons; secondly, a source term which sums  over all source strings and the frequencies  of gravitons produced from them. The integral is weighted by the frequency of the gravitons to obtain their contribution to the energy density. To obtain the spectrum, we write a spectral decomposition
of the graviton energy density at any time $t$ in terms of the graviton angular frequencies today ($\omega_0$): 
\begin{equation}\label{eq:densitylogfreq}
    \rho_{GW}(t)\equiv \int_0^\infty{d \log \omega_0\, \rho_g(\omega_0,t)}\, .\end{equation}
Note that this is different from a spectral decomposition in terms of $\omega(t)$, the graviton frequencies at time $t$.
 Since gravitons free-stream, $\omega(t) = a_0\omega_0 /a(t)$.
The spectral density then satisfies the integral equation
\begin{equation}
\label{s}
    \int_0^\infty{d \log \omega_0 
    \lr{\frac{\partial}{\partial t}+4H}\rho_g(\omega_0,t)}
    = \int_{l_c}^{\infty} \int_0^\infty{\left. \omega^2 \frac{d \Gamma_{o,g}}{d \omega}\right|_{\omega=a_0\omega_0/a(t)} \Tilde{n}_o(l) \, d\log \omega_0 \, dl} \ ,
\end{equation}
where we have changed variables in the right-hand side.
Equating the integrands in \eqref{s} yields
\begin{equation}
\label{eq:evofin}
    \frac{\partial \rho_g}{\partial t}+4H\rho_g= \int_{l_c}^{\infty} { \omega (t)^2\frac{d \Gamma_{o,g}}{d \omega}\big|_{\omega (t)} \Tilde{n}_o(l) \,  \, dl} \  .
\end{equation}

We now write the graviton emission rate by long open strings \eqref{eq:rate} in the general form
\begin{equation}\label{eq:gw-emission-general}
\frac{d \Gamma_{o,g}}{d \omega}= A \lr{\frac{M_s}{M_p}}^2 M_sl (\omega/T_H)^B \sigma(\omega/T_H) 
\frac{e^{-\omega/T_H}}{1-e^{-\omega/T_H}}\, ,
\end{equation}
where $\sigma(x)$ is a greybody factor written with the convention $\sigma (x \to 0)\to 1$, which fixes $A$.
We  work with this general form so that model dependence and various corrections (such as $\alpha'$ corrections, effects of fluxes and non-trivial curvature of the Calabi-Yau, or corrections due to the Hubble expansion) can be incorporated\footnote{We will see later that this form can also incorporate the emission rate given by \cite{Firrotta:2024fvi}.}. 
Interestingly, we find that the results for the spectrum are universal as long as the decay rate falls off exponentially at large $\omega$, i.e., $\sigma(x \to \infty)\to x^c, \, c \in \mathbb{R}$. Recall that we argued in section \ref{sec:GW-spectrum} that
this exponential fall-off is expected to hold in general.
Finally, note that \eqref{eq:gw-emission-general} reduces to the decay rate \eqref{eq:rate} for $B=2$ and greybody factor 
$\sigma(x)=x (1-e^{-x})/4(1-e^{-x/2})^2$, $A=4\Tilde{A}$.\\

An important feature of \eqref{eq:gw-emission-general}
is that the decay rate is proportional to the 
length of the source string (in keeping with the expectation that graviton 
emission from long strings is a local process). This
implies that the $l$ integral\footnote{We note here that, had the decay rate had any other $l$-dependence, this $l$-integral would result in a different power of $L$ than $\rho_b\sim L^2$. If so, the difference (up to order one factors) from~\eqref{eq:gw-spectrum} leads to a different power of $\omega_0/T_H$ and a shift of $B$. The IR properties of the spectrum would change but the position of the peak will not be drastically altered.} in the
right hand side of \eqref{eq:evofin} yields an answer
which is proportional to the total energy density of
the background.
\begin{equation}
\label{back}
\rho_b(t)=M_s^2\int_{lc}^\infty dl'l'\tilde{n}_o(l',t) . 
\end{equation}
Thus, the bath continuously emits a fraction of its energy into GWs; \eqref{eq:evofin}
simplifies to
\begin{equation}
\label{intd}
\frac{\partial \rho_g}{\partial t}+4H\rho_g=A \lr{\frac{M_s}{M_p}}^2 \lr{\frac{T_H^2}{M_s}}
\lr{\frac{\omega}{T_H}}^{B+2}\sigma (\omega/T_H)
\frac{e^{-\omega/T_H}}{1-e^{-\omega/T_H}}
\rho_b \, .
\end{equation}
Furthermore, by making use of the fact that the background energy density evolves as matter
during the Hagedorn phase, i.e., $\rho_b(t) \propto a(t)^{-3}$ we integrate to obtain the spectral function today
\begin{equation}
\label{eq:gws1}
\rho_g^{(0)}=\lr{\frac{a_*}{a_0}}^3 A\,  \lr{\frac{M_s}{M_p}}^2 \rho_b(t_*) \frac{T_H^2}{M_s} \frac{\omega_0}{T_H} \int_{t_s}^{t_{end}}{dt' \lr{\frac{\omega_0}{T_H}\frac{a_0}{a(t')}}^{B+1}
\sigma \lr{\frac{\omega_0}{T_H}\frac{a_0}{a(t')}} 
\frac{e^{-\frac{a_0}{a(t')}\frac{\omega_0}{T_H}}}{1-e^{-\frac{a_0}{a(t')}\frac{\omega_0}{T_H}}}} \, ,
\end{equation}
where $t_s$ is the time corresponding to  the start of the Hagedorn epoch, $t_{\rm end}$ its end (approximated by $LM_s\sim 1$), and $t_*\in (t_s,t_{\rm end})$ is a fiducial time during the epoch. 

It is easiest to understand the spectrum by changing to the dimensionless integration variable to $x\equiv \omega_0 a_0/T_H a(t)$. We finally have
\begin{equation}\label{eq:gw-spectrum}
\rho_g^{(0)}=
\lr{\frac{a_*}{a_0}}^{3/2}
\sqrt{3\rho_b(t_*)}
A\lr{\frac{M_s}{M_p}} T_H^2\, 
\lr{\frac{\omega_0}{T_H}}^{5/2} 
\int_{\frac{\omega_0 a_0}{T_H a_{end}}}^{\frac{\omega_0 a_0}{T_H a_s}}
{dx \, x^{B-3/2}\, \sigma(x)\frac{e^{-x}}{1-e^{-x}}}
 \, .
\end{equation}

We can now see how to modify \eqref{eq:gw-spectrum} to accommodate the findings of \cite{Firrotta:2024fvi}. 
Assuming that gravition emission by open strings is similar to that by closed strings, \cite{Firrotta:2024fvi} modifies \eqref{eq:gw-emission-general} by changing the greybody factor, taking $B=4$, and multiplying by an additional factor of $M_s l$.
The effect is to replace $\rho_b$ in \eqref{intd} with $\rho_b^{3/2}/N_D M_s^2$, which introduces an additional factor of $(a_*/a(t'))^{3/2}$ to the time integral. That is equivalent to taking $B\to B+3/2$ and multiplying the prefactor of \eqref{eq:gw-spectrum} by
$$
 \sqrt{\rho_b(t_*) {a^{3}(t_*) \over a^3_0}} {1 \over {N_D M_s^{2}}} 
\left( {T_H \over \omega_0}\right)^{3/2} 
$$

\subsection{Features of the spectrum}\label{sec:features}

Next, let us analyse the spectrum. To aid the reader, we summarise the results first. The position of the peak \textit{does not} depend on the local string scale, but is essentially determined by the cosmology following the Hagedorn phase -- for standard cosmological evolution
after the epoch the peak is at CMB frequencies. On the other hand, the overall strength of the amplitude is set by the
local string scale, depending linearly on the ratio ${M_{s} \big{/} M_{p}}$. To see these features, let us use the conservation of entropy during the Hagedorn epoch to write $a_s =a_{\rm end}(L_{\rm end}/L_s)^{2/3}$, where $L_s=L(t_s)$ and similarly for $L_{\rm end}$.  Also write $a_{\rm end}T_H=a_0T_0 G X$, with $T_0$ the temperature of the CMB. Here $G\equiv \lr{g_{*,0}/g_{*, {\rm end}}}^{1/3}$ measures the number of thermalized relativistic degrees of freedom at the end of the phase, and  $X$  parameterises the effects of entropy injections, deviations from standard cosmology, and uncertainties associated with the reheating epoch at the end of the Hagedorn phase. We take $a_*=a_{\rm end}$ as our fiducial time, and use $\rho (t_{\rm end})= N_D^2 L_{\rm end}^2 M_s^6$.
For convenience, we also define $\lambda\equiv (15\sqrt{3}A/\pi^2) (M_s/T_H)^2$ and the dimensionless frequency $Y\equiv \omega_0/T_0 GX=2\pi f_0/T_0 GX$.
Finally, we obtain
\begin{equation}
\label{GWspec}
h^2\Omega_{\tti{GW}}= 
\lambda \, 
h^2\Omega_\gamma N_D L_{\rm end} M_s (G X)^4
 \lr{\frac{M_s}{M_p}}\,
Y^{5/2}I\lr{Y,B,\frac{L_s}{L_{\rm end}}} \, ,
\end{equation}
where
\begin{equation}
\label{eq:idefine}
I\lr{Y,B,\frac{L_s}{L_{\rm end}}}=\int_{Y}^{Y(L_s/L_{\rm end})^{2/3}}
{dx\, x^{B-3/2}\, \sigma (x) \frac{e^{-x}}{1-e^{-x}}}\, .
\end{equation}

\begin{figure}
    \centering
    \includegraphics[width=0.9\linewidth]{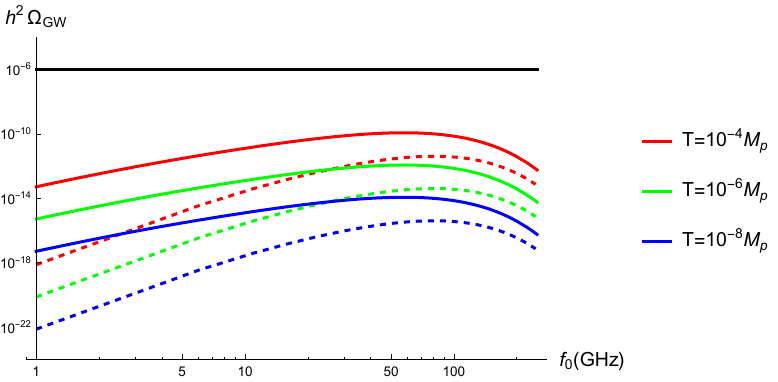}
    \caption{The solid curves show GW spectra for the Hagedorn phase followed by standard cosmology with different values $T_H=\frac{\Upsilon M_s}{2\pi\sqrt{2}}=T$, where heavy open strings radiate gravitons. The reference values taken are $N_D=5$, $L_{\textrm{end}}M_s=5$, $G=0.32$, $X=1$, $A=2\sqrt{2}/(2\pi)^6$, $\Upsilon=1$, $L_s=100L_{\textrm{end}}$. $B$ is taken to be $2$, and $\sigma$ is taken to be $\sigma_{BB}$ given in App.~\ref{app:ShapeSpecDens}. The dashed curves show GW spectra for the SM with different reheating temperatures $T$. The black horizontal line at $10^{-6}$ is a rough bound based on BBN (see footnote~\ref{ft:neff}). The spectra peak at $f_0 \sim 60$ GHz for the Hagedorn phase and at $f_0 \sim 80$ GHz for the SM. Both the axes are taken to be logarithmic.}
    \label{fig:HagedornPhaseVsSM}
\end{figure}

Plots of the spectrum are presented in Fig.~\ref{fig:HagedornPhaseVsSM}. We note that the results
are robust, the basic features are independent of $\sigma(x)$, $L_{\rm end}/L_s$ and $B$, as detailed in App.~\ref{app:robustness} (see, Fig.~\ref{fig:robustness}).
Note that, assuming standard cosmology following the Hagedorn phase, fiducial values render an amplitude\footnote{\label{ft:neff}Recall that Big Bang Nucleosynthesis (BBN) roughly puts $h^2\Omega \leq 10^{-6}$. This is based
on the fact that a GW spectrum of this amplitude and width of the order of the peak would contribute to $\Delta N_{\rm eff}$ at the order one level and BBN sets bounds on
$\Delta N_{\rm eff}$. We will discuss $\Delta N_{\rm eff}$ in the context of our spectrum in detail in section \ref{sec:dr}.}
larger than the typical expectation from the reheating epoch of the Standard Model or Beyond the Standard Model (BSM) theories which were studied in \cite{Ghiglieri:2015nfa,  Ghiglieri:2020mhm, Ringwald:2020ist,Muia:2023wru}.
The remainder of this section illustrates all these points in detail. \\

We note in passing that the effect of using the emission rate of \cite{Firrotta:2024fvi} is to shift $B$ as discussed above and multiply the prefactor by
$M_s L_{\rm end} Y^{-3/2}$ (as well as modifying the greybody factor).

\subsubsection*{Peak amplitude}
We will take $ T_H = \Upsilon {M_s}\big{/}{ 2\pi\sqrt{2} }$ ($\Upsilon =1$ corresponds to the value the Hagedorn temperature in \cite{Kawamoto:2013fza}). Expressing the result in terms of fiducial values one finds\footnote{In principle $g_{*,{\rm end}}$ can be written in terms of $N_D$ and $L_{\rm end}$ in a model-dependent fashion, but we will treat them as separate parameters.}:
\begin{eqnarray}
\label{eq:fidutial-gw-spectrum}
h^2\Omega_{\tti{GW}} &=& 6 \cdot 10^{-11} 
\lr{\frac{N_D}{5}} \lr{\frac{L_{\rm end} M_s}{5}} \lr{\frac{A}{5\cdot 10^{-5} \Upsilon^3}} \frac{\Upsilon}{1} \lr{\frac{G}{0.32} \frac{X}{1}}^{4}
\lr{\frac{M_s}{10^{15} \,\text{GeV}}}
 \cr
\phantom{h^2\Omega_{\tti{GW}}}
&\times&
\lr{\frac{Y}{3}}^{5/2} \frac{I\left(Y,B,L_s/L_{\rm end}\right)}{I\left(3,2,100\right)}\, .
\end{eqnarray}
 
The fiducial value for the dimensionless frequency is taken at $Y = 3$, which is a frequency of around 60 GHz (the actual peak frequency is somewhat larger and so is the associated amplitude - we are quoting the approximate value for $B=2$).
To be more concrete about the position of this peak, note that for $Y \gg 1$ the amplitude is exponentially suppressed while for $Y \ll 1$ it grows polynomially.
It must, therefore, admit a peak when $Y \simeq 1$ (quantitative statements depend on $B$ and $\sigma (x)$ but the qualitative behaviour is the same under our assumptions), which, assuming standard cosmology, correlates the peak frequency with the CMB temperature.
We will study deviations from standard cosmology below.

\subsubsection*{Behaviour far from the peak}
Gravitational wave astronomy is being developed for a wide range of frequencies, and it is interesting to obtain as many features as possible from a given source. Even though the signal will be faint away from the peak, it is noteworthy that the spectral index is not too large, and so the amplitude is meaningful for a range of frequencies.
Let us examine its behaviour at low frequencies.
The starting point is the $Y$-dependent part of Eq.~\eqref{GWspec}.
We observe the following regimes:
\begin{itemize}
    \item Ultra-low frequencies: $Y\ll \lr{L_{\tti{end}}/L_s}^{2/3}$.
    
    In this case the integral can be well approximated by (recalling by definition $\sigma (x\to 0)\to 1$):
    \begin{equation}
       \Omega_{\tti{GW}} \sim Y^{5/2}I\lr{Y\ll \lr{\frac{L_{\tti{end}}}{L_s}}^{2/3},B, \frac{L_s}{L_{\tti{end}}}}\simeq 
        \frac{Y^{B+1}}{B-3/2}\lr{\lr{\frac{L_s}{L_{\tti{end}}}}^{2B/3-1}-1}  
    \end{equation}
    The amplitude grows like $f_0^{B+1}$.
    \item Intermediate frequencies $1>Y> \lr{L_{\tti{end}}/L_s}^{2/3}$.

    In this case the behaviour is qualitatively different depending on $B$.
    To make progress, let us fix $\sigma (x)=1$ noting that the qualitative changes encoded in $\sigma (x)$ will not significantly alter our conclusions.
    The idea is then that for $B \geq 5/2$ the integrand has a saddle and the integral is well approximated by a constant.
    Otherwise the behaviour is as above.
    Thus:
    \begin{equation}
    \Omega_{\tti{GW}} \sim
        \begin{cases}
        Y^{5/2} \, , \qquad (B > 5/2)\, , \\
        Y^{B+1} \, , \qquad (B< 5/2)
        \end{cases}
    \end{equation}
    The $f_0^{B+1}$ behaviour is reproduced if $B$ is small enough.
    Otherwise we observe an universal $f_0^{5/2}$ behaviour.
    This is to be interpreted as a superposition of the peak signals from early times.
    \item High frequencies $Y \gg 1$.

    The amplitude is exponentially suppressed.
\end{itemize}

\subsubsection*{Dependence of the spectrum on the posterior cosmology}\label{sec:posterior-cosmology}
We can study effects of the posterior cosmology in the shape of the gravitational wave background following the discussion in~\cite{Muia:2023wru}.
These effects are encoded in the parameter $X$. It is  easy to see that variations in $X$ $(X \to X')$ can be incorporated by considering  $ \omega_0 \to \omega_0 X'/X$ and $\Omega \to\Omega (X'/X)^4$. 
Typical epochs that contribute non-trivially to X are those that
lead to injection of entropy due to degrees of freedom becoming massive, and those involving deviations from radiation domination (such
as epochs of modulus domination). See \cite{Muia:2023wru} for details.\\

We can also use the help of the parameter $X$ to be as conservative as we may want with regard to how much we can trust the last moments of the Hagedorn phase.
Instead of setting $L_{\tti{end}}M_s=1$, we could set $L_{\tti{end}}M_s=10$ (for which the large $N$ limit in Eq.~\eqref{eq:prob-ratio-G} holds appropriately~\cite{Frey:2023khe}; this arises from the exponential behaviour of the density of states being a good approximation).
If so, noting that conservation of entropy requires $L \sim a^{-3/2}$, this would reduce $X$ by a factor $10^{2/3}\sim 5$.
The peak frequency is thus redshifted to around half its value, and the peak amplitude (taking into account the overall $L_{\tti{end}}$ pre-factor) diminishes by a factor $10^{5/3}\sim 50$.\\

In general, the peak frequency behaves like $f \sim L_{\tti{end}}^{-2/3}$ and the peak amplitude satisfies $\Omega \sim L_{\tti{end}}^{-5/3}$.
We thus conclude that the general features of the spectrum are very robust: a growing spectrum that peaks at the GHz band which is largely independent of the inherent compactification, the duration and details of the Hagedorn phase, and even of its ending.
The largest uncertainties arise from the details of the cosmology after the phase.

\subsection{Bounds on dark radiation}
\label{sec:dr}

       There are bounds on the total energy density in any form of dark radiation  from the observations of the cosmic microwave background
and big bang nucleosynthesis \cite{Planck:2018nkj, Cyburt:2015mya}. These apply to stochastic gravitational wave backgrounds.
In our setting, the
energy density in gravitational waves can be obtained by integrating $\rho_g^{(0)}/\omega_0$ in \eqref{eq:gws1} over $\omega_0$. The required integrations can be performed by carrying out variable change
$$
  \omega' = { \omega_0 a_0 \over T_{H} a(t) } \ \ \ \ \textrm{and} \ \ \ \  t' = t  ;
$$
and performing the $\omega'$ integration first. This yields
\begin{equation}
\label{e11}
\rho_{\rm GW}^{(0)} = \mathcal{I} {{A }T^2_{H} \over M_s}\left( {M_s \over M_p} \right)^2 \left({ a(t_{*}) \over a(t_0)}\right)^3 \rho_{b}(t_{*}) 
\int_{t_s}^{t_{\rm end}} dt \left( { a(t) \over a(t_0)} \right)
\end{equation}
where  $\mathcal{I}=\int_0^{\infty } \frac{x^{B+1} \sigma (x)e^{-x}}{1-e^{-x}} \, dx$.\footnote{Note that for $B=2$, when $\sigma=1$ (i.e., in absence of greybody factor), or when $\sigma=\sigma_{BB}$ or $\sigma_{FF}$ as given in App \ref{app:ShapeSpecDens}, $\mathcal{I}$ takes the following values, respectively: $\frac{\pi^{4}}{15},\frac{32}{15} \left(\pi ^4-90 \zeta (5)\right),\frac{16}{3} \left(9 \zeta (3)-\pi ^2\right)\approx 6.49, 8.72, 5.06$. These values increase monotonically as we increase $B$. For example, for $B=3$, the respective values are $\mathcal{I}\approx 24.87, 37.60, 17.47$.}
Finally,  changing the integration variable  in \eqref{e11} to the scale factor one obtains:
\begin{eqnarray}
\label{gsimplit}
\rho_{\rm GW}^{(0)}
&=& \sqrt{3}\mathcal{I} {{A }T^2_{H} }\left( {M_s \over M_p} \right) 
\left[\left({ a_{*} \over a_0}\right)^4 \rho_{b}(t_{*}) \right]{1 \over \sqrt{\rho{(t_{*})}}} {1 \over a_{*}^{5/2}}
\int_{a_s}^{a_{\rm end}} d a a^{3/2}
\end{eqnarray}
It is simplest to take $t_{*} = t_{\rm end}$, the integral
in \eqref{gsimplit} is then dominated by its upper limit and one has
\begin{equation}
\label{efor}
   \rho_{\rm GW}^{(0)} \approx {2 \sqrt{3} \over 5} \mathcal{I} {{A }T^2_{H}
   \over\sqrt{\rho{(t_{\rm end})}}}\left( {M_s \over M_p} \right) 
\left[\left({ a_{\rm end} \over a_0}\right)^4 \rho_{b}(t_{\rm end}) \right]
\end{equation}
 Note that our answer is independent of $L$ i.e how close the universe was to the Hagedorn temperature at  the onset of
 the epoch.

    The above allows us to carry out an important consistency
check on our calculations. We have worked assuming that 
during the Hagedorn epoch, the energy density in the gravitons is much less than that of the background (this was implicit
in our calculations, we took the background to be evolving like matter during the epoch).  Note that the above formula for the energy density in gravitational waves applies for any time $t_0$ which
is greater than or equal to $t_{\rm end}$ ($t_0$ need not 
correspond to today). Taking it to be $t_{\rm end}$ the consistency condition translates to
\begin{equation}
\label{cccc}
 {2 \sqrt{3} \over 5} \mathcal{I} {{A }T^2_{H}
   \over\sqrt{\rho{(t_{\rm end})}}}\left( {M_s \over M_p} \right) \ll 1  \  \ \
   \textrm{i.e.} \ \ \    \mathcal{I}A {\sqrt{3} \over 20 \pi^2 } { \Upsilon^2 \over {{L_{\rm end}} M_s N_D}}\left( {M_s \over M_p} \right) \ll 1
\end{equation}
Recall that $\rho(t_{\rm end})=  N_D^2 L_{\rm end}^2 M_{s}^6$, its the square root is  greater that $T_{H}^2$. Thus the condition \eqref{cccc} is milder than the requirement that effective field theory is
valid at the onset of the Hagedorn phase $(N_D L M_{s} { M_{s} \over M_p} \ll 1, \text{Eq. \ref{eq:hubble2}})$ and
is automatically satisfied if the effective field theory
is valid.

The abundance in dark radiation is typically reported
in terms of $\Delta N_{\rm eff}$ (the effective number of additional neutrino like species at the time of neutrino decoupling). This can be related to energy densities at
the time of reheating by the  formula \cite{Cicoli:2012aq}:
\begin{equation}
\label{drfor}
    \Delta N_{\rm eff}  = {43 \over 7} { \rho_{\rm dark} \over \rho_{\rm vis} } \left( { g(T_{\nu}) \over g(T_{\rm rh}) } \right)^{1/3}
\end{equation}
Thus, if one takes the entire energy density in the background to decay
to the visible sector, making use of \eqref{drfor} and \eqref{cccc} one has 
\begin{equation}
\label{Nbound}
\Delta N_{\rm eff} \approx {{43 \sqrt{3}} \over {140 \pi^2}}A \mathcal{I} \left( { g(T_{\nu}) \over g(T_{\rm rh}) } \right)^{1/3}    { \Upsilon^2 \over {{L_{\rm end}} M_s N_D}}\left( {M_s \over M_p} \right) 
\lesssim \mathcal{O} \lr{{M_s \over  M_p}}
\end{equation}
The careful reader might have noticed the difference in the
parametric dependence on $N_{D}$ between \eqref{GWspec} and \eqref{Nbound}. The reason for this as follows: in \eqref{GWspec}
the energy is the visible sector was taken to be as 
given by observations (as per equation \eqref{phenoGW}); on the other
hand, in \eqref{Nbound} the energy density in the visible sector
at the end of the Hagedorn epoch is set to be equal the entire
energy density in the background. Of course, for a model to be
successful the visible sector energy density at the end
of the epoch should evolve to what is observed. In such a setting,
$N_{D}$ would not be a free parameter but determined by that the consistency
condition set by  this evolution (in addition to the requirement that
all the Standard Model degrees of freedom are realised).
Since we do not carry out any model building related to the visible sector in this paper, we do not include any factors of 
$N_{D}$ in the bound set in \eqref{Nbound}.\\

Present observational bounds put $\Delta N_{\rm eff} < 0.2$ 
(see e.g. \cite{Planck:2018nkj, Cyburt:2015mya}) which cannot be reached in our setup if computational control is preserved (which we recall required a hierarchy $M_s \ll M_p$).
However, noting that future experiments (see e.g. \cite{CMB-S4:2016ple, SimonsObservatory:2018koc}) will  
probe $\Delta N_{\rm eff}$ at much smaller values, our considerations could be used to bound the string scale in these scenarios.
Furthermore, the condition for the validity of effective field theory
\eqref{eq:hubble2} can be used to put an upper
bound on $\Delta N_{\rm eff}$:
$$
\Delta N_{\rm eff} < \mathcal{O} \lr{{ 1 \over N^2_{D} L_{\rm start}}}
$$

\subsection{Comparison with the Standard Model }

Next, let us turn to a comparison of our results with a stochastic gravitational wave background that is produced
from a reheating epoch in field theory.
It is well understood~\cite{Ghiglieri:2015nfa,Ghiglieri:2020mhm,Ringwald:2020ist,Muia:2023wru} that a thermal bath sources out of equilibrium gravitons (as in our case), and that the GW spectrum today is dominated by emission at the earliest times.
This is opposite to our case because the gas of strings behaves nonrelativistically and so the energy density in GWs sourced earlier is redshifted away.\\

It is true in both cases, however, that the energy density in gravitons is proportional to the characteristic scale with Planckian suppression: $T/M_p$ in field theory, $M_s/M_p$ in string theory.
That is, the amplitude grows linearly with the reheating temperature.
It is also true in both cases (assuming standard cosmology after emission) that the peak frequencies lie around $50-100$ GHz and an amplitude linear in the reheating temperature, as illustrated in Fig.~\ref{fig:HagedornPhaseVsSM}.
Indeed, one may think of both spectra as having the same origin: the GWs arising from a thermal phase in the early Universe.\\

Our computation thus fixes the high-energy, stringy part of the spectrum and its amplitude turns out to be larger.
The reason is that the leading-order string process occurs at 3-points (is a decay), which -when allowed- is typically subdominant in field theory\footnote{To see this, notice that the decay should be suppressed by $m/M_p$ for a particle of mass $m$ decaying through gravity-mediation.
If $m \leq T$, the effect is negligible compared to $T/M_p$, and otherwise the number of particles decaying is exponentially suppressed.}.
The leading contribution in field theory involves four external legs instead~\cite{Ghiglieri:2015nfa,Ghiglieri:2020mhm} and is therefore suppressed by higher powers of couplings.
It is worth remarking~\cite{Ringwald:2020ist,Muia:2023wru} that in absence of exotic physics after reheating, the amplitude at a given reheating temperature is not expected to be parametrically larger than the Standard Model prediction.
The Hagedorn phase overcomes this conclusion due to the stringy feature of an exponentially growing density of states (which allows for states with masses larger than the temperature to be excited).\\

It is worth remarking that we trust our computations whenever this exponential is a good approximation of the density of states, that is, $L \gtrsim l_c$.
There is an intermediate regime before standard reheating in which some massive degrees of freedom are excited and source GWs.
We expect that an intermediate spectrum is sourced in this regime which interpolates between the high-energy string behaviour described here and the low-energy field theoretical computations in~\cite{Ghiglieri:2015nfa,Ghiglieri:2020mhm}.

\section{Conclusions and future directions}
\label{sec:conclude}

In this paper, we have considered the early universe at high temperatures with long open and closed strings
in thermal equilibrium (at temperatures close to the Hagedorn temperature). The equilibrium energy density is dominated by
open strings, the strings are non-relativistic and the Universe evolves as a matter-dominated era.
The universe evolves adiabatically --- the temperature falls and the average length
of the string decreases. The epoch ends with the strings decaying primarily to massless open string
degrees of freedom (which correspond to the Standard Model degrees of freedom).
We have studied the stochastic gravitational wave background produced as a result
of graviton emission by long open strings during such an epoch\footnote{As discussed in section \ref{sec:thcosmo}, the conditions needed for
validity of our analysis are compatible with inflation --- i.e. it is consistent
to have such an epoch after inflation leading to observable consequences (if inflation did take place and there were such epochs before it, there would be 
no observable consequences due to dilution).}. The two main features of the spectrum
are:
\begin{itemize}
   \item The location of the peak
is essentially determined by post-inflationary history. For standard
cosmological evolution after the epoch, the peak is close to the peak
of the CMB.
  \item The amplitude does not depend strongly on the details
  of the epoch. The magnitude is stronger than what is expected from
  the reheating epoch of the Standard Model or BSM models for string-scale reheating.
\end{itemize}

The present work opens up several interesting avenues. Here, we have focused
on the production of gravitons during the Hagedorn phase.
It is important to examine other cosmological consequences of
such an epoch,
e.g., the production of KK gravitons and their connection with
dark matter (following \cite{Kofman:2005yz, Frey:2009qb}),  emission of bulk axions
\footnote{A possible concern is whether the production of light, long wavelength modes (e.g. Kahler moduli) could destabilize the compactification. We do not expect this to occur because the energy deposited on these modes is a very small fraction of the energy of the plasma (as is the case for gravitons). In addition, their production is incoherent so we do not expect them to contribute to the semiclassical background configuration. Further analysis is left for future work.} which leads to the production of dark radiation (and implications for entropic arguments
for suppression in the abundance of dark radiation resulting from Hagedorn phase \cite{Frey:2021jyo}) and connections with non-standard cosmological histories (recent string
cosmology reviews --
\cite{Cicoli:2023opf, Brandenberger:2023ver} and references therein provide
detailed descriptions of such possibilities). Also, highly excited strings have many properties of quantum black holes (see e.g.
\cite{tHooft:1990fkf, Susskind:1993ws, Amati:1999fv}). The phenomenology of the universe whose
constituents are produced from the decay of primordial black holes (via Hawking
radiation) has been explored in \cite{Lennon:2017tqq} and related works. A comparative study will be interesting.\\

The exponentially growing string density of states is a key
input for our calculations and is one of the factors that
distinguishes the setting from a field-theoretic setting.
Epochs of cosmological stasis
\cite{Dienes:2021woi,Dienes:2022zgd,Dienes:2023ziv}
also feature a tower of states.  It will be interesting
to study the stochastic gravitational wave background
produced during such cosmological epochs and compare it with our results.\\

This paper has not touched upon the question, ``What conditions in the early universe lead to an epoch of Hagedorn cosmology?'' (only the associated energetics). The end of brane-antibrane inflation at the bottom of a warped throat is natural setting \cite{Frey:2005jk,Frey:2009qb}. Another possibility is
at the end of modulus-mediated kination (see \cite{Conlon:2022pnx,Apers:2022cyl,Apers:2024ffe,Conlon:2024uob} for a recent discussion of kination in string
cosmology). It is important to develop a better understanding of this question.
Also, as noted, the open string Hagedorn phase acts like a matter dominated stage of cosmic evolution.
Consequences of early matter domination, including effects on dark matter abundance and evolution of cosmological perturbations, have been studied extensively in the context of cosmological moduli; see \cite{Kane:2015jia} for a review. 
Understanding if any of these effects differ or whether the small pressure ($\propto\sqrt{\rho}M_s^2$) has significant effects are important.
Further, it may be interesting to revisit our calculation if there are other contributions to the energy density beyond the long strings.\\

On the more formal side, it will be interesting to revisit graviton emission from long strings
using the methods of \cite{Firrotta:2024fvi} and compare
with the results of \cite{Amati:1999fv}. It will also be interesting to study the decay rate
in realistic compactifications by making  use of the 
string field theoretic methods developed in \cite{Cho:2023mhw} (see~\cite{Sen:2024nfd} for a recent review), and compare our results with more general considerations~\cite{Sen:2024bax}.
Many of these directions are already under study.\\

We would like to end by emphasising an important point tied to the fact that the amplitude of the gravitational wave produced depends linearly on $M_{s} \big{/} M_{p}$ i.e. the signal  {\it grows} with increased $M_s$. On the other hand, signals associated with the supersymmetry breaking scale/ KK modes probed in particle colliders diminish with increased $M_{s}$ 
(this is because  mass scales such as soft masses and the mass of KK modes, scale as a positive power of $M_{s} \big{/} M_{p}$, and the signal decreases as
these scales increase).  Thus, the gravitational wave background potentially
provides a new probe of the string scale precisely in regimes where it would be difficult to access using more traditional probes.\\

Furthermore, in contrast to these low-energy processes that, if observed, may be explained by non-stringy EFT models, our calculations show that the spectrum we found is string theoretical in nature and cannot be reproduced by the usual field theoretic extensions of the Standard Model in thermal equilibrium\footnote{It remains to be seen whether any other physics may be able to reproduce the features of the spectrum we found, including the IR spectral index.}. This illustrates the UV sensitivity of gravitational wave backgrounds and gives additional motivation  towards future efforts on gravitational wave astronomy \cite{Aggarwal:2020olq}.

\section*{Acknowledgements}
We would like to especially thank Francesco Muia for very enjoyable earlier collaborations and useful discussions. We also acknowledge useful conversations with  Steve Abel, Santiago Agui, Fien Apers, Michele Cicoli, Joe Conlon, Sebastian Cespedes, Shanta de Alwis, Maurizio Firrotta, Chris Hughes, Elias Kiritsis, Mario Ramos-Hamud, Filippo Revello,  and Jorge Russo.  This work  is partly based upon work from COST Action COSMIC WISPers CA21106, supported by COST (European Cooperation in Science and Technology); AM, RM, FQ and GV are participants of the COST Action. FQ and GV acknowledge the CERN Theory Department for hospitality.
The work of FQ and GV has been partially supported by STFC consolidated grant ST/T000694/1 and ST/X000664/1.
The work of AF and RM has been supported by the Natural Sciences and Engineering Research Council of Canada Discovery Grant program, grant 2020-00054.

\appendix

\section{String thermodynamics in realistic setups}
\label{sec:setup}
\subsection{Limiting and non-limiting string thermodynamics}
We begin the discussion by pointing out that the canonical ensemble can only describe string thermodynamics in certain situations, and we would like to understand whether realistic scenarios with string-size energy densities admit this description.
The point is that the density of states $d(l)$\footnote{We use string units $M_s=1$ in this subsection, so $l$ is a measure of the energy of the string.} generically reads~\cite{Brandenberger:1988aj,Deo:1988jj,Deo:1989bv,Deo:1991mp,Abel:1999rq,Mitchell:1987hr,Mitchell:1987th}
\begin{equation}
    d(l)\sim l ^{-A}e^{\beta_H l}\, .
\end{equation}
Whenever the canonical ensemble describes the thermodynamics, $n(l)\sim d(l)e^{-\beta l}$,
so, depending on the value of $A$, the total energy density 
\begin{equation}
\rho\propto\int_{l_c}^{\infty}{dl \, l^{-A+1}e^{-l/L}}\, 
\end{equation} 
may diverge or not as the Hagedorn temperature ($L\to \infty )$ is approached. Note that all compact dimensions are small in this limit because long strings fill a linear scale $L_{rms}=\sqrt{L}\to\infty$.

The case where $\rho$ is finite in this limit can be understood as leading to a phase transition at this critical energy density (which is order one in string units).
This behaviour was called non-limiting in~\cite{Abel:1999rq}, in the sense that the Hagedorn temperature can be achieved with finite energy.
In the case of closed strings only in three or more compact dimensions, most of the energy of the high-energy phase is in a very few long strings \cite{Mitchell:1987hr,Mitchell:1987th,Deo:1989bv,Abel:1999rq,Barbon:2004dd,Copeland:1998na} (if in fact there is not gravitational collapse to black holes).

For $A\leq 2$, $\rho$ diverges as $L\to\infty$, which is known as the limiting case. The canonical ensemble is valid for large energy densities\footnote{until the energy density is large enough $\rho \sim 1/g_s$ to nucleate brane-antibrane pairs.}.
In this case, there is a continuous change from a radiation (massless string) gas to a gas of highly excited strings as the energy density and temperature increase. This occurs for open (and closed) strings whenever no more than four of the dimensions transverse to the D-branes are noncompact.\\

In cosmology, we expect the initial string gas to have a large but finite energy density (for example in reheating after inflation), so we are most interested in the thermodynamics with finite $L\leq L_s$.
As described at length in~\cite{Abel:1999rq,Frey:2023khe}, the number $A$ depends on the ratio of the typical string extent $L_{rms}=\sqrt{L}$ with respect to a set of characteristic lengths in the compactification.
The idea (which is well understood in a Boltzmann equation approach) is that the equilibrium distributions are determined by the decay rates, and these are themselves weighted by the probability that a string self-intersects (or can be chopped by a brane).

Let us begin by describing the case of closed strings.
A typical highly excited string in flat space will form a random walk, with typical spread $l^{1/2}$.
In $D$ spatial dimensions, the string will therefore fill a $D$-dimensional cube of volume $l^{D/2}$, and so the probability\footnote{Strictly speaking this is the probability that an open random walk self-intersects, but the qualitative argument applies equally well for closed random walks.} that two points of the string separated by $l$ will self-intersect is $1/l^{D/2}$.
If instead the string is placed in a $D-d$-dimensional cube of side $a \ll l^{1/2}$, the string will fill up the whole box and , 
so the probability will be $1/(l^{d/2}a^{D-d})$ instead.
This is illustrated in Fig.~\ref{sf:large-small}.
We would say, in the latter case, that the string perceives $d$ large directions and $D-d$ small directions.
To make this more concrete, for typical string length $L$\footnote{this is not generally the average string length but is the same up to an order one factor}, when $\sqrt{L}\gg a$, the gas of strings would perceives $d$ large directions, and $D$ otherwise.

The case of open strings is qualitatively different but can be understood in similar terms.
The difference is that an open string need not self-intersect in order to decay --- it just needs to get in contact with a brane.
Consider a D$p$-brane.
An open string with its endpoints in this brane can decay if it is touching the brane.
That is, the point of the string needs to come back in the directions orthogonal to the brane.
In this case we should therefore only look at the $D-p$ directions orthogonal to the brane.
Again, letting these orthogonal directions be a $(D-p)$-dimensional box of length $a$, a string of size $l$ with $l^{1/2}\gg a$ will perceive $d_\perp=0$ large orthogonal directions, and $d_\perp=D-p$ if $l^{1/2}\ll a$ instead.

The case is again different if we include a homogeneous gas of parallel D$p$-branes along $q$ of the orthogonal directions.
This introduces a new scale: the inter-brane separation, $l_b$.
If the size of the string satisfies $l^{1/2}\gg l_b$, the string does not need to come back to its original brane in order to decay: for semiclassical matters, the branes are perceived by the highly excited string as \textit{effectively overlapping} and (again, for decay matters), the gas of branes in these directions behaves like an extension of the worldvolume.
This is illustrated in Fig.~\ref{sf:gas-branes}.

Having understood the definitions of $d$ and $d_\perp$, we can now write the equilibrium distributions in general~\cite{Abel:1999rq,Frey:2023khe}.
These read, for open and closed strings respectively,
\begin{equation}
    n_o(l) \sim l^{-d_\perp/2}e^{-l/L} \, , \qquad n_c(l) \sim l^{-1-d/2}e^{-l/L}\, .
\end{equation}
Thus, canonical ensemble methods apply whenever $d_\perp \leq 4$ or\footnote{If several systems are put in contact, the overall system will feature limiting behaviour if one of its parts does~\cite{Abel:1999rq}.} $d\leq 2$.
The possibilities are summarized in table~\ref{tab:limiting}.
\begin{table}[h!] 
\centering
\begin{tabular}{ |c||c|c|c|c|  }
 \hline
Condition  & Closed ($d$) & Open ($d_\perp$) & Limiting & Without open \\
 \hline\hline
 $l_b \gg L_{rms}\gg l_s$ & $9 $   & $9-p$ &   $p\geq 5$ & No\\
 \hline
 $l_{KK} \gg L_{rms} \gg l_b$ &  $9$  & $9-p-q$   & $p+q\geq 5$ & No\\
 \hline
 $l_{U}\gg L_{rms} \gg l_{KK}$ & $3$ & $0$ &  Yes & No\\
 \hline
 $ L_{rms} \gg l_{U}$ & $0$ & $0$ &  Yes & Yes \\
 \hline
\end{tabular}
\caption{Limiting vs non-limiting behaviour for open and closed strings in 3+6 dimensions.
The 6 (3) dimensions are assumed isotropic and span a length $l_{KK}$ ($l_U$).
We also allow for the presence of a gas of space-filling parallel D$p$-branes in $q$ directions.}
\label{tab:limiting}
\end{table}

\subsection{Three possible scenarios}

The discussion above allows us to identify three possible scenarios that would render $d_\perp=0$, which is the setup that we will consider in this paper.
This allows for a Hagedorn phase with highly excited open and closed strings for a large range of energy densities.
Noting that, as we will see, the dominant contribution to the GW spectrum arises from the later times, the cases that can accommodate the lowest energy densities are the most interesting ones. 
Reviewing from section \ref{sec:scenarios}, the three possibilities are
\begin{itemize}
    \item  {\it Open String Brandenberger-Vafa scenario~\cite{Brandenberger:1988aj}}.
    It is a possibility that the typical strings were much longer than the Kaluza-Klein scale $l_{KK}$, in a way similar in spirit to the endpoint of the Brandenberger-Vafa scenario, where the $3$ large spatial directions have decompactified. 
    In this case, $\Omega_\perp$ (and $\Omega_\|$ if the branes extend in the compact dimensions) are large, which can suppress equilibration rates in a model-dependent manner.
    As the temperature decreases, this is followed by a stage of long strings with $d_\perp>0$, which we do not consider in this manuscript (it is at lower density and therefore should contribute less to gravitational waves).
    \item {\it Dense brane scenario}.
    If there is a homogeneous distribution of branes along all directions in the compact space, then the strings only need to be as large as the typical inter-brane separation which would therefore work at energies lower than the Brandenberger-Vafa case.
    In this case, potentially all of the dimensions are large, so the $b_1$ terms in the Boltzmann equations \eqref{eq:boltzmann1} are modified to
$(b_1/2N_d)(\tilde{n}_o(l)/l^{(3+d)/2}$, where $d$ is the number of large compact dimensions. 
The coefficent $b_1\simeq g_s\Omega_\perp^l/M_s^{(1+d)/2}\Omega_\|^s$, where $s$ and $l$ indicate the size of small and large dimensions respectively.
We consider the cosmology and gravitational wave emission of the corresponding thermodynamics in future work.
As the universe cools, the system becomes a gas of long strings on isolated branes (with large transverse dimensions); again, we do not consider this stage of evolution.
    \item {\it  Jackson-Jones-Polchinski~\cite{Jackson:2004zg} box scenario}.
    As argued originally in \cite{Jackson:2004zg}, backreaction of fluxes and other ingredients needed for moduli stabilization generates a potential that localises highly excited strings (and branes) in a volume of order 1 in string units.
    Because the strings are localised at the minimum of the warp factor, the 4D EFT includes long strings at the warped string scale $M_s$\footnote{In general the warped string scale may be replaced by the so-called moduli dependent species scale, assuming the light tower of states is an effective string \cite{Dvali:2007hz,Ooguri:2006in}.}.
    In this case the thermodynamic description works down to typical lengths of order $L\sim 1/M_s$, and the gas continuously changes to standard radiation with SM fields.
    This is the most interesting case from the perspective of a realistic compactification.
\end{itemize}

In this article we will assume the last case, which is the most conceivable because it takes into account the effect of ingredients present generically in realistic setups. 

\subsection{Review of the JJP box}
Let us thus review~\cite{Jackson:2004zg}, which considered warping (while neglecting possible effects due to flux).
Note that strong warping such as a long throat is not necessary; order 1 fluctuations of the warp factor suffice to localize the strings.

The idea is to show that the wavefunction of a long string in a warped background is localized in a (fundamental) string-sized box.
To do so, we wish to find the probability of fundamental strings to intersect in a warped background. In a warped region, long strings are attracted to the bottom of a warped throat due to a worldsheet potential.\footnote{$(p,q)$ strings likewise feel a potential due to variations of the dilaton.} 
To see this more clearly, write the worldsheet action\begin{equation}
S=-\frac{1}{2\pi \alpha'}\int{d^2 \sigma \, \lr{-\text{det} (h_{ab})}^{1/2}}\, ,
\end{equation}
in a warped background
\begin{equation}
h_{ab}=e^{2A(Y)} \eta_{\mu \nu} \partial_a X^\mu \partial_b X^\nu + e^{-2A(Y)}g_{ij}(Y) \partial_a Y^i \partial_b Y^j\, .
\end{equation}
In a static gauge with $\sigma$ identified with the string's extent in $X^\mu$, we find a potential
\begin{equation}
V(Y)=\frac{e^{2A(Y)}}{2\pi \alpha'} ,
\end{equation}
so the action to second order in $Y$ (choosing coordinates so that the minimum of the potential lies at $Y=0$ and $e^{-2A(0)} g_{ij}(0)=\delta_{ij}$) is
\begin{equation}
S \simeq -\int{d^2 \sigma \,  \lr{V(0)+\frac{1}{2}\partial_i \partial_j V(0) Y^i Y^j+\frac{1}{4\pi \alpha'}\partial_a Y^i \partial^a Y^i }}\, ,
\end{equation}
ie, the fluctuations around the minimum are massive worldsheet scalars. 
At zero temperature, the two-point correlator for each worldsheet scalar $Y^i$ (at a single specified worldsheet position) is an integral and sum (with cut off) over wavenumbers $(k^0,k^1=2\pi n/l')$\footnote{assuming closed strings for simplicity; the open string case is similar.} 
\begin{equation}\label{eq:correlator}
\langle Y^i Y^i\rangle =
\frac{\alpha'}{l'}\sum_{n} \int \frac{dk^0}{(k^0)^2+(2\pi n/l')^2+V_{,ii}(0)}
\to\frac{\alpha'}{2\pi}\int^\Lambda \frac{d^2k}{k^2+V_{,ii}(0)} ,
\end{equation}
in the long string limit, where $\langle Y^2\rangle$ independent of the string length.
Here, $l'$ is the length of the string projected in the $X^\mu$ directions, which is an order 1 factor times the total length $l$ of the string for typical configurations. The initial prefactor of $1/l'$ follows from normalization of the modes of $Y^i$ on the finite spatial extent of the string.

The natural cut-off of the worldsheet theory with $X^0$ as the time coordinate is the warped string scale, so
we find a logarithmic correlator
\begin{equation}
\langle Y^i Y^i\rangle =\frac{\alpha'}{2}\log \lr{1+\frac{1}{2\pi \alpha' e^{-2A(0)}V_{,ii}(0)}}\equiv \frac{\alpha'}{2}\omega_i\, 
\end{equation}
Near the Hagedorn temperature (measured in warped units), the $k^0$ integral becomes a sum over Matsubara frequencies (with cutoff) with the effect that $\omega_i\to 2\pi/\beta_H\sqrt{V_{,ii}}$ \cite{Frey:2005jk}, which is parametrically similar in the limit of slowly varying warp factor. 
The key point is that two-point function is independent of the length of the string in the long string limit both at $\beta_H$ and zero temperature.

The intersection probability, or overlap of the string positions, 
is therefore where
the worldsheets of coordinate fluctuations $Y$ and $Y'$ coincide:
\begin{equation}\label{eq:coupling-long}
\langle \delta^6 (Y-Y') \rangle= 
\int{\frac{d^6k}{(2\pi)^6} \, \langle e^{i k \cdot (Y-Y')} \rangle }= 
\int{\frac{d^6k}{(2\pi)^6} \, e^{-k_i k_j \langle (Y-Y')^i (Y-Y')^j \rangle /2} } 
\simeq \frac{1}{\alpha'^3\Pi_{i=1}^6 \omega_i^{1/2}}\, ,
\end{equation}
assuming that all the strings are confined in the same place. Since we can treat the spacetime positions of different points on the worldsheet of a long string, this also applies to the self-intersection probability of a single long string.\\

Since $\omega_i$ is typically order unity in a warped throat, we find that the string interaction rates behave as if the strings fill a compact space that is a bit larger than string scale.\footnote{Strictly speaking, the bottom of a prototypical Klebanov-Strassler warped throat \cite{Klebanov:2000hb} is $\mathbb{R}^3\times S^3$ with the warp factor independent of the $S^3$ directions; we assume that the thermal fluctuations are large enough to fill the somewhat-larger than string scale $S^3$.}
Critically, this intersection probability is independent of the length of each string. A classical interpretation is that it is proportional to the geometric probability of one point on each string occupying the same point in a 6D compact space of volume $\alpha'^3\Pi_{i=1}^6 \omega_i^{1/2}$.

One may similarly compute the interaction between a string and a $D$-brane at location $Y_D$ (assumed fixed) to be
\begin{equation}\label{eq:coupling-brane}
\langle \delta^6 (Y-Y_D) \rangle
\simeq \frac{1}{\alpha'^3\Pi_{i=1}^6 \omega_i^{1/2}}e^{-\sum_{i=1}^6 \frac{Y^i_D Y^i_D}{\alpha' \omega_i}}\, .
\end{equation}
If the strings and brane are localised at different points in the extra dimensions, the overlap of their wavefunctions is significantly small, reducing the interaction probability.
Henceforth, we will assume that the gas of long strings is attracted to the position in the compactification of the D-branes.\\

We are now in position to argue that string thermodynamics in a warped compactification is described similarly to that in 3 noncompact and 6 small compact dimensions without warping.
Comparing with the discussion at the beginning of the section, we see that the interaction rates have the same scaling as those for long strings with effectively compact dimensions (ie, dimensions filled by the random walk of a typical length string); namely, they are inversely proportional to a constant compact volume --- in this case $\alpha'^3(\prod \omega_i)^{1/2}$.
For example, consider the rate for a closed string of length $l$ to split on a stack of $N_D$ D3-branes also localized at the tip of a warped throat.
From (\ref{eq:coupling-brane}), this rate should be $\Gamma\propto g_s N_D l/(\prod\omega_i)^{1/2}$, where the factor of $l$ accounts for the number of points on the string that can intersect the brane ((\ref{eq:coupling-brane}) is the probability for a specified point on the string to intersect the brane).

As a result, we approximate the thermodynamics of strings in a warped compactification by strings in flat spacetime with a compactification of volume $\sim(\prod\omega_i)^{1/2}$ in string units; the long strings fill those compact dimensions.
We can also approximate the warp factor as constant across the compact region (its curvature should be small in string units, and $\omega_i\sim 1$).
Then the string thermodynamics are those of strings in three flat noncompact dimensions with spacefilling D-branes; as argued in \cite{Frey:2023khe}, the effect of the compact dimensions is just to modify the interaction coefficients.
Also, since the 4D metric $g_{\mu\nu}$ appears in the 10D line element as $ds^2 =e^{2A(y)} g_{\mu\nu}dx^\mu dx^\nu+\cdots$, the energy scale of the strings is set by the warped string scale, which we denote $M_s$.\\

Finally, we note that these arguments do not apply to massless string states, including gravitons. Since these strings are not long, we take the sum from \eqref{eq:correlator}. 
The $n=0$ term is proportional to $1/l'$, which indicates that light strings can probe beyond the quadratic term in the Taylor expansion of $V(Y)$. 
As a result, construction of massless vertex operators requires considering the entire CFT.
We use the 10D supergravity description to determine the profile of the massless string zero modes.

\section{Aspects of the gravitational wave spectrum }
\label{aspects}

  In this appendix, we will study various aspects of the gravitational wave spectrum in detail. As in the rest of the paper, we will
set up the analysis relying on a characterisation of the decay rates  based on $\sigma(x)$ and B (as introduced in Eq. \ref{eq:gw-emission-general}) and specialise to particular forms
to extract specific features.

\subsection{The shape of the spectral density}
\label{app:ShapeSpecDens}

  In this section we will compare the shape of the spectrum of gravitational waves in our
  scenario to that of a black body. From \eqref{GWspec}, one sees that the general
  form of the spectrum of gravitational waves (per unit frequency interval as opposed to unit logarithmic frequency interval) is:
  \begin{equation}
  \label{phenospec}
  f(\omega_{0}) = \alpha (\omega_0 \hat{T})^{3/2} \int_{{\omega_{0}/ \hat{T}}}^{{R\omega_{0}/ \hat{T}}} dx { \sigma(x) x^{B-3/2} \over {e^{x} - 1} }\,,
\end{equation}
where $\alpha, R, \hat{T}$ are determined in terms of the details of the compactification,
the temperature at the onset of the Hagedorn phase and details of the reheating phase.
$\sigma(x)$ is the greybody factor of the microscopic decay rate.
In this subsection, we  will treat $\alpha, R, \hat{T}$  as phenomenological parameters. Note that
$\alpha, R$ are dimensionless, while $\hat{T}$ has dimensions of temperature. Also,
$R$ sets the duration of the Hagedorn phase, our interest lies in $R \gg 1$.\\

In order to compare the shape of the spectrum to the blackbody shape, we will choose the parameters such that the total energy density and peak of the spectrum match with a blackbody (at temperature $T$) and then plot both the spectra in the same graph. We will take $B=2$.\\

Let us consider the cases:

\begin{equation}
    \sigma(x)=1\,, \qquad 
    \sigma_{BB}(x)=\frac{x \left(e^x-1\right)}{4 \left(e^{x/2}-1\right)^2}\,, \qquad     \sigma_{FF}(x)=\frac{4 \left(e^x-1\right)}{x \left(e^{x/2}+1\right)^2}\,, 
\end{equation}
normalised such that $\sigma(x)\to1$ as $x\to0$. The last two are the greybody factors respectively for emission of an NS-NS massless boson and an R-R massless boson from a typical open superstring \cite{Kawamoto:2013fza}. $BB$ denotes an emitted closed string state with bosonic oscillators on both left and right moving parts, while $FF$ denotes that with fermionic oscillators. The respective total energy densities associated with the spectrum \eqref{phenospec} in limit $R \gg 1$ are\footnote{These are obtained by first changing the integration variable in \eqref{phenospec} to $y \equiv x {\hat{T} \over \omega_{0}}$, then carrying out the integral over $\omega_0$ and finally carrying out the integral of $y$.}

\begin{equation}
    \hat{\rho} = \alpha {2 \over 5} {\pi^{4} \over 15} \hat{T}^{4}\,, \qquad 
    \hat{\rho}_{BB}=\frac{64}{75} \alpha \left(\pi ^4-90 \zeta (5)\right) \hat{T}^4\,, \qquad 
    \hat{\rho}_{FF}=\frac{32}{15} \alpha \left(9 \zeta (3)-\pi ^2\right) \hat{T}^4\,. 
\end{equation}
The peaks of respective spectra can be determined numerically as:

\begin{equation}
    \omega_0 \simeq 1.66 \hat{T}\,, \qquad 
    \omega^{BB}_0\simeq 1.87 \hat{T}\,, \qquad
    \omega^{FF}_0\simeq 1.52 \hat{T}\,. 
\end{equation}

\begin{figure}
    \centering
    \includegraphics[width=0.9\linewidth]{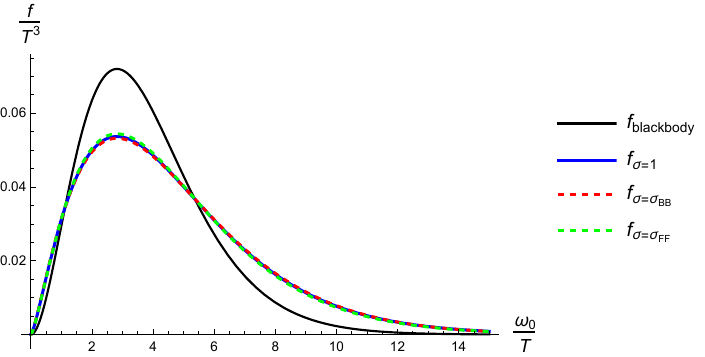}
    \caption{A comparative plot which contrasts our spectra \eqref{phenospec} with that of the blackbody spectrum. Here, the parameters of our spectra are chosen so that the peak and total energy density are same as those of a blackbody spectrum at temperature $T$. Note that our spectra have a smaller value at maximum and are much broader than the blackbody spectrum.}
    \label{fig:gw-spectrum}
\end{figure}

Recall that for the black body spectrum at temperature $T$ (for degeneracy factor g=1)
\begin{align}
  f_{\rm blackbody}(w_0) = {1 \over 2 \pi^{2}} { w_0^{3} \over {e^{w_0/T} - 1}}\,,
\end{align}
the total energy density and the location of the peak are respectively
\begin{align}
  \rho_{\rm blackbody} = {1 \over 2 \pi^{2}} {\pi^{4} \over 15} T^{4}\,,\quad w_0^{\rm blackbody} \simeq 2.82 T\,.
\end{align}
Thus, in each of the three cases above requiring that a spectrum of the form in \eqref{phenospec} has the same peak and total energy density as those of a blackbody at temperature $T$ yields $\alpha$ and relates $\hat{T}$ linearly to $T$:
\begin{align}
  \sigma=1:\qquad&\hat{T}=1.69 T\,,\quad \alpha= 0.015\,;\nonumber\\
  \sigma=\sigma_{BB}:\qquad&\hat{T}=1.51 T\,,\quad \alpha= 0.018\,;\\
  \sigma=\sigma_{FF}:\qquad&\hat{T}=1.86 T\,,\quad \alpha= 0.014\,.\nonumber\\
\end{align}
We plot the spectrum \eqref{phenospec} with these parameters alongside a blackbody spectrum at temperature $T$ in Fig. \ref{fig:gw-spectrum}.

\subsection{Robustness under $\sigma,\ B,\ L_s/L_{\textrm{end}}$}
\label{app:robustness}

\begin{figure}
    \centering
    \includegraphics[width=1.02\linewidth]{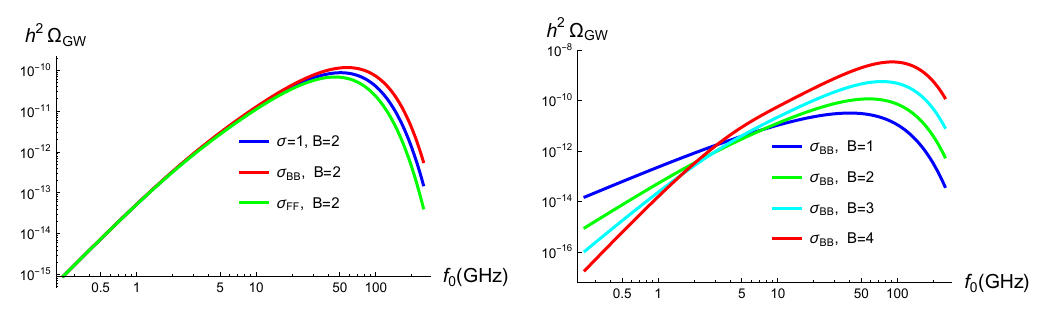}
    \caption{GW spectra for the Hagedorn phase with $T_H=\frac{\Upsilon M_s}{2\pi\sqrt{2}}=10^{-4}M_p$. In the left panel, different $\sigma$ given in appendix \ref{app:ShapeSpecDens} have been considered with $B=2$ that corresponds to 4D flat space background. In the right panel, different $B$ values have been considered for $\sigma=\sigma_{BB}$. Reference values for other parameters are taken as: $N_D=5$, $L_{\textrm{end}}M_s=5$, $G=0.32$, $X=1$, $A=2\sqrt{2}/(2\pi)^6$, $\Upsilon=1$, $L_s=100L_{\textrm{end}}$. All axes are taken to be logarithmic.}
    \label{fig:robustness}
\end{figure}

The gravitational wave spectrum \eqref{GWspec} is robust under $\sigma,\ B,\ L_s/L_{\textrm{end}}$. For example, the dependence on $L_s/L_{\textrm{end}}$ comes through a factor $I\lr{Y,B,\frac{L_s}{L_{\rm end}}}$. For large $L_s/L_{\textrm{end}}$, $I$ can be well approximated by $\int_{Y}^{\infty}
{dx\, x^{B-3/2}\, \sigma (x)\frac{1}{e^x-1}}$ which is clearly independent of $L_s/L_{\textrm{end}}$. Assuming $\sigma\sim x^C$ for large $x$, the correction term goes as $e^{-Y(\frac{L_s}{L_{\textrm{end}}})^{2/3}} (\frac{L_s}{L_{\textrm{end}}})^{\frac{2}{3}(B+C)-1}$ which is exponentially suppressed. The spectrum has been plotted against different $\sigma(x)$ and $B$ values in Fig.~\ref{fig:robustness}.
\newpage
\bibliography{biblio}
\bibliographystyle{utphys}

\end{document}